\documentclass[aps,pre,twocolumn,noshowpacs,nofootinbib,superscriptaddress]{revtex4-1}

\usepackage{amsmath}
\usepackage{amsfonts}
\usepackage{amssymb}
\usepackage{amsthm}
\usepackage{mathrsfs}
\usepackage{dsfont}
\usepackage{esint}
\usepackage{graphicx}
\usepackage{physics}

\DeclareMathAlphabet{\mathpzc}{OT1}{pzc}{m}{it}

	\newcommand{\mr}[1]{\mathrm{#1}}			

	\newcommand{\br}[1]{\left( #1 \right)}
	\newcommand{\brr}[1]{\left[ #1 \right]}
	
	\newcommand{\of}[1]{\!\br{#1}}
	\newcommand{\off}[1]{\!\brr{#1}}
	
	\newcommand{\sbr}[1]{( #1 )}
	\newcommand{\sbrr}[1]{[ #1 ]}
	\newcommand{\sbrrr}[1]{\{ #1 \}}
	\newcommand{\sof}[1]{\!\sbr{#1}}
	
	\newcommand{\sofff}[1]{\!\sbrrr{#1}}

	\newcommand{\Sum}[2]{\sum\limits_{#1}^{#2}}

	\newcommand{\Int}[3]{\int\limits_{#1}^{#2}\mr{d}#3\,}
	
	\newcommand{\sSum}[2]{\sum_{#1}^{#2}}

	\newcommand{\Landau}[1]{\mathpzc{O}\of{#1}}
	\newcommand{\sLandau}[1]{\mathpzc{O}\sof{#1}}

		\newcommand{\Id}{\mathds{1}}

		\newcommand{\Span}[1]{\mathrm{Span}\off{#1}}

\newcommand{\Ham}{\hat{H}}
\newcommand{\TEO}{\hat{U}}

\newcommand{\Detect}{\hat{D}}
\newcommand{\PsiDet}{\psi_\text{d}}
\newcommand{\RDet}{r_\text{d}}
\newcommand{\PsiIn}{\psi_\text{in}}
\newcommand{\RIn}{r_\text{in}}
\newcommand{\FDA}{\varphi}

\newcommand{\SP}{S}
\newcommand{\TDP}{P_\text{det}}

\newcommand{\Surf}{\hat{\mathfrak{S}}}

\newcommand{\AUS}{u}

\newcommand{\Hilbert}{\mathcal{H}}
\newcommand{\PBright}{\hat{P}_{\Hilbert_B}}

\begin{document}

\title{Dark states of quantum search cause imperfect detection}

\author{Felix Thiel}
\affiliation{Department of Physics, Institute of Nanotechnology and Advanced Materials, Bar-Ilan University, Ramat-Gan 52900, Israel}
\email{thiel@posteo.de}
\author{Itay Mualem}
\affiliation{Department of Physics, Institute of Nanotechnology and Advanced Materials, Bar-Ilan University, Ramat-Gan 52900, Israel}
\author{Dror Meidan}
\affiliation{Department of Physics, Institute of Nanotechnology and Advanced Materials, Bar-Ilan University, Ramat-Gan 52900, Israel}
\author{Eli Barkai}
\affiliation{Department of Physics, Institute of Nanotechnology and Advanced Materials, Bar-Ilan University, Ramat-Gan 52900, Israel}
\author{David A. Kessler}
\affiliation{Department of Physics, Institute of Nanotechnology and Advanced Materials, Bar-Ilan University, Ramat-Gan 52900, Israel}

\begin{abstract}
  We consider a quantum walk  where a detector repeatedly probes the system  with fixed rate $1/\tau$
  until the walker is detected. 
  This is a quantum version of the first-passage problem.
  We focus on the total probability, $P_{\mathrm{det}}$, that the particle is eventually detected in some target state, for 
  example on a node $r_{\mathrm{d}}$ on a graph, after an arbitrary number of detection attempts.
 Analyzing the dark and bright states for finite graphs, and more generally for systems with a discrete spectrum, we provide an explicit formula for $P_{\mathrm{det}}$ in terms of the energy eigenstates which is generically $\tau$ independent. 
We  find that disorder in the underlying Hamiltonian renders perfect detection:  $P_{\mathrm{det}}=1$, and then expose
the   role of symmetry with respect to sub-optimal detection. Specifically  we give a simple upper bound for $P_{\mathrm{det}}$ that is controlled by 
the number of equivalent (with respect to the detection) states in the system. 
We also extend our results to infinite systems, for example the detection probability of a quantum walk on a line, which is $\tau$-dependent and less than half, well below Polya's
optimal detection for a classical random walk.
\end{abstract}

\maketitle
\section{Introduction}
  
  Recently the first detection problem for quantum dynamics has attracted increasing interest \cite{Bach2004a, Krovi2006a, Krovi2006b, Krovi2007a, Stefanak2008a, Varbanov2008a, Caruso2009a, Agliari2010a, Gruenbaum2013a, Bourgain2014a, Krapivsky2014a, Dhar2015a, Dhar2015b, Sinkovicz2015a, Sinkovicz2016a, Lahiri2019a, Friedman2017a, Friedman2017b, Thiel2018a, Thiel2018b, Thiel2020a, Thiel2020c, Thiel2020d%
  },  due in part to its potential relevance  for the readout of certain quantum computations.
  More fundamentally, it sheds light on hitting time processes and measurement theory in quantum theory \cite{Gurvitz2017a, Ashida2018a, Buffoni2019a, Chan2019a, Skinner2019a}. 
  The classical counterpart of this topic is the first passage time problem, 
  which has a vast number of applications in many fields of science \cite{Redner2007a, Raposo2009a, Benichou2011a, Palyulin2016a, Godec2016a, Godec2016b}.
  In its simplest guise, a classical random  walker initially located on a particular vertex of a graph is considered, 
  and the question of interest is: When will the particle arrive at a target state, for instance another
  vertex of the graph, for the first time?
  For  the quantum system, we investigate unitary evolution on a graph, with a particle in an initial 
  state $\ket{\PsiIn}$, which could be, e.g., a localized state on a vertex $\ket{\RIn}$.
  This evolution is repeatedly perturbed by detection attempts for another state $\ket{\PsiDet}$ 
  called the detection state (for example, another localized vertex state of the system $\ket{\RDet}$, see below).  
  In this situation, the concept of first arrival is not meaningful, but we can register the 
  first {\em detected} arrival time.
  The protocol of measurement (i.e. the epochs of the detection attempts) crucially determines this first detection time \cite{Gruenbaum2013a,Friedman2017b}.
  We consider a stroboscopic protocol, i.e., a sequence of identical measurements with fixed inter-attempt time $\tau$, continued 
  until the first successful detection (see definitions below).
  One of the  general aims in this direction of research is to gain information on the statistics of this event,
  which is inherently random by the basic laws of quantum mechanics. 
  Our approach explicitly incorporates repeated, strong measurements into the definition of the first detected arrival, 
  and therein differs from other quantum search setups \cite{%
    Grover1997a, Aaronson2003a, Bach2004a, Childs2004a, Muelken2006a, Perets2008a, Karski2009a, %
    Zaehringer2010a, Magniez2011a, Muelken2011a, Jackson2012a, Novo2015a, Chakraborty2016a, %
    Boettcher2015a, Preiss2015a, Xue2015a, Li2017a, Mukherjee2018a, Rose2018a%
  } and from the time-of-arrival problem \cite{%
    Allcock1969a, Kijowski1974a, Aharonov1998a, Damborenea2002a, Anastopoulos2006a, Halliwell2009a,%
    Ruschhaupt2009a, Sombillo2014a, Sombillo2016a%
  }.

  In some cases, quantum search is by far more efficient than its classical counterpart. 
  In particular for the hypercube and for certain trees it was shown that quantum search can be 
  exponentially faster than  possible classically \cite{Farhi1998a, Childs2002a, Kempe2005a, Krovi2006a}.
  Indeed, while the classical random walker repeatedly re-samples its trajectory, a quantum walker may benefit
  from the constructive interference of its wave function.
  This mechanism enables a quantum walker to achieve much faster detection times than his classical counterpart.
  In the same way, however, certain initial conditions suffer from destructive interference, such that
  the desired state is never detected and yields a diverging mean detection time \cite{Krovi2006a, Krovi2006b, Krovi2007a, Varbanov2008a, Friedman2017b}.
  We call such initial conditions {\em dark states}.
  The terminology {\em dark states} is borrowed from atomic physics and quantum optics 
  where it describes forbidden transitions or non-emissive states \cite{Plenio1998a,
Stefani2009a}.
  The classical random  walk, if the process is ergodic, does not possess dark states and hence in this sense performs ``better'', since detection on a finite graph is guaranteed. 
  In the quantum problem, the presence of the detector splits up the total Hilbert space into a dark space \cite{Krovi2006a,Facchi2003a,Caruso2009a} and its complement.
  These play the roles of the ergodic components in a classical random walk.
  In contrast to the classical situation, they are not generated by a separation of state space (alone), but rather by destructive interference.

  The main focus of this paper is the total detection probability $\TDP$.
  This is the probability to detect the particle eventually, namely the
detection probability in principle after an infinite  number of attempts
(though the measurement process is stopped once the particle is detected). 
In a finite system, if the initial and detection states coincide, $\TDP$ is always unity~\cite{Gruenbaum2013a}. However, when the initial state differs from the detection state, the initial state can have an overlap with some dark states, which are undetectable,
  and the overlap of the initial state with the dark space gives no contribution to the
  $\TDP$.
 We derive an explicit formula for $\TDP$ in terms of the eigenstates $\ket{E_{l,m}}$ of the unitary propagator $\TEO(\tau) := e^{-i\tau \Ham/\hbar}$ (or, the Hamiltonian $\Ham$), the $m$th state  with quasienergy $\lambda_l = \tau E_l/\hbar \mod 2\pi$, such that $\TEO(\tau)\ket{E_{l,m}} = e^{-i\lambda_l}\ket{E_{l,m}}$:
   \begin{equation}
    \TDP(\PsiIn)
    =
    \sideset{}{'} \sum_l
    \frac{
      \abs*{
        \sSum{m=1}{g_l} \ip*{\PsiDet}{E_{l,m}} \ip*{E_{l,m}}{\PsiIn}
      }^2
    }{
      \sSum{m=1}{g_l} \abs*{\ip*{E_{l,m}}{\PsiDet}}^2
    }
    .
  \label{eq:Kessler}
  \end{equation}
This is the first main result of this manuscript.
  For most values of $\tau$, there is a one-to-one correspondence between the set of values of $E_l$ and those of $\lambda_l$. Only, if there are one or more pairs of energies $\{E_k,E_l\}$, such that the resonance condition 
  \begin{equation}
  (E_k-E_l)\tau/\hbar = 0 \mod 2\pi
  \label{eq:DefResonantTau}
  \end{equation}
  is satisfied are there fewer quasienergies than energies.
  The primed sum runs over those distinct quasienergy sectors that have non-zero overlap with the detected state, (i.e., excluding any completely dark sectors) with the inner sum running over the $g_l$  degenerate states of quasienergy level $\lambda_l$.  Thus, except for the zero-measure set of resonant values of $\tau$, $\TDP$ does not depend on the detection period at all. However, at these resonant $\tau$s, $\TDP$ changes dramatically \cite{Yin2019a}.
  This formula is obviously invariant under a change of basis within any quasienergy sector. We give two different derivations of this formula, one based on the
quantum renewal formula~\cite{Gruenbaum2013a,Friedman2017b} for the probability of detection after $n$ measurements, and a more elementary proof  based on an explicit formula for the bright and dark states in terms of the propagator's eigenstates. This proof requires showing that, on a finite graph, all states that are orthogonal to the dark space are bright, that is, they are detected with probability one, and so $\TDP$  is equal to the initial state's overlap with the bright space. Both our renewal formula derivation and this latter proof break down in the cases of infinite graphs, where the spectrum has a continuous part. We will discuss this point in detail in the context of the infinite line.

\begin{figure}
    \centering
    \includegraphics[width=0.99\columnwidth]{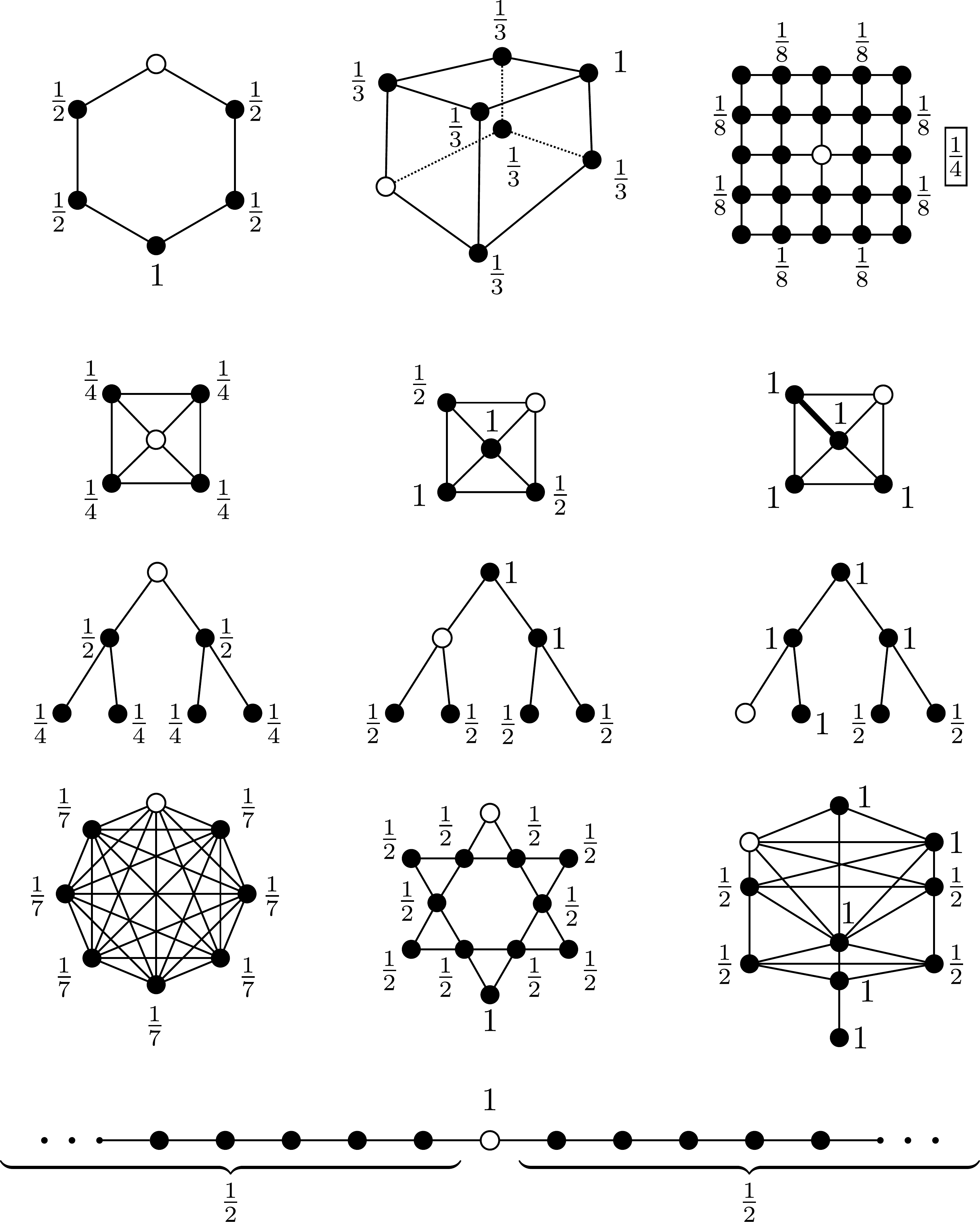}
    \caption{
 The upper bound for some simple graphs showing the deviations of detection
probability $P_{{\rm det}}$
from the classical counter part which is unity.
      The numbers represent the bound for $\TDP$ from Eq.~\eqref{eq:UpperBound}.
      An open circle denotes the detection site $\RDet$, and any other node is a possible localized initial state $\ket{\PsiIn} = \ket{\RIn}$.
      The quantum particle resides on the nodes of these graphs and travels along its links which are all identical ($\hat{H}$ is the adjacency matrix of the graph).
      In all graphs, the on-site energies are equal to zero.
      From left to right and top to bottom:
      The ring of size six, the hypercube of dimension three, a two dimensional simple cubic lattice,
      the square graph with detection site in the center, in a corner, and in a corner with with one modified link,
      the binary tree graph in two generations with detection in the root, middle and leaves,
      the complete graph with eight sites, the Star-of-David graph, and the Tree-of-Life graph.
      The infinite line is shown at the very bottom.
      The numbers are upper bounds for $\TDP$ which are easy to obtain,
 they will be later compared with the exact result
 Eq.~(\ref{eq:Kessler}).
    }
    \label{fig:PD}
  \end{figure}

   While Eq.~\eqref{eq:Kessler} is  the exact solution to the problem, insight is found addressing the symmetry of $H$ and for systems whose Hamiltonian is
given by an adjacency matrix the symmetry of  the graph. 
 Krovi and Brun 
    \cite{Krovi2006a}
already showed that the total detection probability is sub-optimal, or in their language the hitting
time is infinite, if the system exhibits symmetry. 
 Indeed as indicated by  
 Eq.~\eqref{eq:DefResonantTau}
 degeneracy plays an important role in the evaluation of $P_{{\rm det}}$
and  hence 
symmetry of the underlying unitary  is crucial. 
In the second part of this paper,  we obtain a remarkably simple upper bound $P_{{\rm det}}\le 1/\nu$ where $\nu$ is the number of distinct sites of the graph
equivalent to the initial localized initial state. 
Briefly,
 that means that  we search for  $\nu$ graph nodes
that are equivalent with respect to the detector  (which is also localized on a node) and this number yields the mentioned bound on the detection
probability, see Fig. \ref{fig:PD} for details. 
More general symmetry consideration will follow.

  The rest of this paper is organized as follows:
  In Sec.~\ref{sec:Strobo} we will introduce our model.
  Then we derive our main result, first using the renewal formula in Sec.~\ref{sec:Cauchy}. After that we  
discuss 
  the splitting of the Hilbert space into bright and dark subspaces  in Sec.~\ref{sec:Dichotomy} and use this in Sec.~\ref{sec:TDP2} to re-derive our main formula.
  Examples are discussed in Sec.~\ref{sec:Examples}.  The case of the infinite line is analyzed in Sec.~\ref{sec:Line}. 
 Finally we provide an upper bound exploiting the  symmetry of the underlying graph. 
 A brief summary of some of our results was presented in Ref. \cite{Thiel2020a}.

\section{Stroboscopic detection protocol}
\label{sec:Strobo}
  One does not simply observe the first arrival of a quantum particle in some target state $\ket{\PsiDet}$,
  because it does not have a trajectory in the classical sense.
  The measurement, the detection, must be explicitly incorporated into the dynamics.
  Following Refs.~\cite{Gruenbaum2013a, Bourgain2014a, Dhar2015a, Dhar2015b, Friedman2017a, %
  Friedman2017b, Lahiri2019a, Thiel2018a, Thiel2018b} this can be done by adhering to the 
  {\em stroboscopic detection protocol}, where detection in state $\ket{\PsiDet}$ is attempted at the times 
  $\tau, 2 \tau, 3 \tau , \hdots $  and so on.
  The detection period $\tau$ between the detection attempts is a parameter of the experimentalist's choice.
  The experiment we have in mind follows this protocol:
  \begin{itemize}
    \item[1.] 
      Prepare the system in state $\ket{\PsiIn}$ at time $t=0$ and set $n=0$.
    \item[2.] 
      The system evolves unitarily for time $\tau$ with the evolution operator $\TEO(\tau) := e^{-i\tau \Ham/\hbar}$; 
      the wave function is then $\ket{\psi(n\tau+\tau^-)} = \TEO(\tau)\ket{\psi(n\tau)}$.
      [The $-$($+$) superscript denotes the limit from the below (above)].
      Increase $n$ by one, $n=n+1$.
    \item[3.] 
      Attempt to detect the system in the state $\ket{\PsiDet}$ with a strong, collapsing measurement.
      \begin{itemize}
    \item[a.] 
      With conditional probability  
      $\norm*{\Detect\ket{\psi(n\tau^-)}}^2 
      = \abs*{\ip*{\PsiDet}{\psi(n\tau^-)}}^2$,
      the system is successfully detected.
      Here, 
      \begin{equation}
      \Detect = \dyad{\PsiDet}
      \end{equation}
      is the projector onto the detection state.
      The detection time is $t = n \tau$ and the experiment ends.
    \item[b.] 
      Otherwise,  the measurement failed to detect the system in the target state.
      The wave function is instantaneously projected to the state that has no 
      overlap with the detection state $\ket{\PsiDet}$. 
      This is the collapse postulate \cite{Cohen-Tannoudji2009a}.
      Mathematically the wave function directly after the unsuccessful measurement
      is equal to $\ket{\psi(n\tau^+)} = N_n\sbrr{ \Id - \Detect } \ket{\psi(n\tau^-)}$,
      where $N_n$ is a normalization constant, and $\Id$ is the identity operator.
      After constructing the new wave function, jump back to step two.
      This loop is repeated until the system is finally detected in step 3a.
  \end{itemize}
  \end{itemize}
  After following this procedure many times, one may construct a histogram for the first successful 
  detection number $n$.

As shown by Dhar, et al.~\cite{Dhar2015a}, the overall probability of detection at measurement $n$ is
\begin{equation}
F_n =  \norm*{\Detect\TEO(\tau)[(\Id-\Detect)\TEO(\tau)]^{n-1}\ket{\PsiIn}}^2
\end{equation}
 and the probability of no detection in the first $n$ measurements is
  \begin{align}
    S_n(\PsiIn) &= 1- \sum_{m=1}^{n} F_m 
    = \norm*{[(\Id-\Detect)\TEO(\tau)]^n\ket{\PsiIn}}^2  \nonumber\\
    &= \norm*{\Surf^n\ket{\PsiIn}}^2  ,
  \label{eq:Sn}
  \end{align}
  where $\Surf := (\Id-\Detect)\TEO(\tau)$ is the survival operator.
  The dependence on the initial state is stressed in the notation.
  The dependence on the detection state, however, will be suppressed throughout the article.
  Clearly, $S_n$ involves $n$ compound steps of unitary evolution followed by unsuccessful detection.
  The main focus of this paper is the total detection probability,
  the probability to be eventually detected, i.e. the probability to ``not survive'':
  \begin{equation}
    \TDP(\PsiIn)
    = \sum_{n=1}^\infty F_n =
    1 - \lim_{n\to\infty} S_n(\PsiIn)
    .
  \label{eq:DefTDP}
  \end{equation}
  An initial state $\ket{\PsiIn}$ that is never detected is called a {\em dark state} with respect to 
  the detection state $\ket{\PsiDet}$; for these states $\TDP(\PsiIn) = 0$ and $S_n(\PsiIn) = 1$ for all $n$.
  Similarly, a {\em bright state} is detected with probability one; i.e., $\TDP(\PsiIn) = 1$, and $S_n(\PsiIn) \to 0$.
  Of course we may also have states that are neither dark nor bright.

  Our theory is developed in generality, valid for any finite dimensional Hamiltonian $\Ham$.
  Besides the initial and detection state, $\Ham$ and the detection period $\tau$ are the  
  ingredients that enter the stroboscopic detection protocol via the evolution operator $\TEO(\tau) := e^{-i \tau \Ham / \hbar}$.
  We assume that a diagonalization of the latter is available:
  \begin{equation}
    \TEO(\tau)
    =
    \Sum{l}{} e^{- i \lambda_l} \hat{P}_l
    \qc
    \hat{P}_l := \Sum{m=1}{g_l} \dyad{E_{l,m}}
    .
  \label{eq:TEODiag}
  \end{equation}
  Here, $\ket{E_{l,m}}$ are the eigenstates and $\hat{P}_l$ are the eigenspace projectors, 
  that gather all eigenstates of the $g_l$-fold degenerate quasienergy level $\lambda_l=\tau E_l/\hbar \mod 2\pi$, so that all $\lambda_l$ in Eq.~\eqref{eq:TEODiag} are distinct. This form is easily obtained from a similar diagonalization of the Hamiltonian 
  $\Ham = \sSum{l}{} E_l \sSum{m=1}{g_l'} \dyad*{E_{l,m}'}$.  For convenience, we label the states by their energy $E_l$, rather than by their quasienergy $\lambda_l$, since we focus on non-resonant values of $\tau$.

  Although the only effect of $\tau$ on $\TDP$ is these discontinuous changes, $\tau$ has a more profound effect on other quantities like the mean detection time \cite{Friedman2017b, Liu2020a}.
  It is an important parameter of the stroboscopic detection protocol.
  In the limit $\tau\to0$, when the system is observed almost continuously, the Zeno effect freezes 
  the dynamics \cite{Misra1977a, Itano1990a, Elliott2016a, Li2018a} and detection can become impossible.
  
\section{From the renewal equation to the total detection probability}
\label{sec:Cauchy}
  One proof of Eq.~\eqref{eq:Kessler} starts from 
 the first detection amplitudes $\FDA_n$ \cite{Friedman2017a}, which in turn yields the first detection probabilities $F_n=|\FDA_n|^2$, and so $\TDP = \sum_{n=1}^\infty F_n$.
  The generating function for these amplitudes $\FDA(z) = \sSum{n=1}{\infty} z^n \FDA_n$ 
  can be expressed in terms of $\TEO(\tau)$ (for details, see Refs.~\cite{Friedman2017a, Friedman2017b}):
  \begin{equation}
    \FDA(z)
    =
    z
    \frac{ 
      \mel*{\PsiDet}{\frac{\TEO(\tau)}{\Id - z \TEO(\tau)}}{\PsiIn}
    }{
      \ev*{\frac{\Id}{\Id - z \TEO(\tau)}}{\PsiDet}
    }
    .
  \label{eq:DefGenFunc}
  \end{equation}
  This formula is  obtained from the quantum renewal equation: a basic tool for the derivation of the detection amplitude \cite{Gruenbaum2013a,Friedman2017b}. 
  As shown in \cite{Friedman2017b}, $\TDP$ can be obtained from  $\FDA(z)$ :
  \begin{align}
    \TDP
    = &  \nonumber
    \Sum{m,n=1}{\infty} \delta_{m,n} \FDA_m^* \FDA_n
    =
    \frac{1}{2\pi} \Int{0}{2\pi}{\theta}
    \Sum{m,n=1}{\infty} 
    \FDA_m^* \FDA_n e^{i \theta(n-m)}
    \\ = &
    \frac{1}{2\pi} \Int{0}{2\pi}{\theta}
    \abs*{\FDA(e^{i\theta})}^2 ,
  \label{eq:TDPHardy}
  \end{align}
  i.e., as the integral of the generating function's modulus on the unit circle.
  
  Using the quasienergy representation of the evolution operator, \eqref{eq:TEODiag},  the generating function \eqref{eq:DefGenFunc}
  is expressed as the fraction of two expressions:
  \begin{equation}
    \FDA(z) 
    = 
    \frac{
      \sSum{l}{}
        \mel*{\PsiDet}{\hat{P}_l}{\PsiIn}
        \frac{
          z e^{-i \frac{\tau E_l}{\hbar}}
        }{
          1 - z e^{- i \frac{\tau E_l}{\hbar} }
        }
    }{
      \sSum{l}{} \frac{
        \ev*{\hat{P}_l}{\PsiDet}
      }{
        1 - z e^{- i \frac{\tau E_l}{\hbar} }
      }
    }
    =
    z
    \frac{(K \mu \nu )(z)}{(K\mu)(z)}
    .
  \label{eq:FDAGenFunc}
  \end{equation}
  $(K\mu)(z)$ is the so-called Cauchy transform of the function $\mu(\theta)$, $0 \le \theta < 2\pi$, defined by \cite{Cima2006a}:
  \begin{equation}
    (K\mu)(z)
    =
    \frac{1}{2\pi}
    \Int{0}{2\pi}{\theta}
    \frac{\mu(\theta)}{1 - z e^{-i\theta}}
    .
  \label{eq:DefCauchy}
  \end{equation}
  In Ref.~\cite{Thiel2018b}, we showed that the 
   denominator of $\FDA(z)$ is a Cauchy transform of the so-called 
  wrapped measurement spectral density of states:
  \begin{align}
    \mu(\theta)
    := & \nonumber 
    2\pi \ev*{\delta( e^{-i \theta} - \TEO(\tau))}{\PsiDet}
    \\ = &
    2 \pi \Sum{l}{} \ev*{\hat{P}_l}{\PsiDet} \delta(\theta - \lambda_l)
    ,
  \label{eq:Cauchy1}
  \end{align}
  where $\lambda_l$ is the phase corresponding to the $l$-th quasienergy level: $\lambda_l := \tau E_l / \hbar \mod 2\pi$.
  All $\lambda_l$s are distinct.

  The numerator is a Cauchy transform as well, but of a product of functions $\mu(\theta) \nu(\theta)$:
  \begin{align}
    \mu(\theta) \nu(\theta)
    :=
    \sideset{}{'} \sum_l
    \ev*{\hat{P}_l}{\PsiDet}
    \delta(\theta - \lambda_l)
    \underbrace{
      \frac{
        \mel*{\PsiDet}{\hat{P}_l}{\PsiIn} e^{-i\lambda_l}
      }{  
        \ev*{\hat{P}_l}{\PsiDet}
      }
    }_{=: \nu(\lambda_l)}
    .
  \end{align}
  Almost all values of the functions $\nu(\theta)$ are irrelevant, except  at the specific arguments $\lambda_l$,
  where they are fixed, $\nu(\lambda_l) = e^{-i\lambda_l} \mel*{\PsiDet}{\hat{P}_l}{\PsiIn} / \ev*{\hat{P}_l}{\PsiDet}$.
  [Due to the delta functions in $\mu(\theta)$, these are the only significant values of $\nu(\theta)$.]
  The primed sum excludes all completely dark energy levels, i.e., those for which $\hat{P}_l\ket{\PsiDet}=0$, so that $\ev*{\hat{P}_l}{\PsiDet}$ does not vanish in the denominator of $\nu(\theta)$.
  Putting the numerator and denominator together, we find that $\FDA(z) = z (K\mu\nu)(z) / (K\mu)(z) =: z (\mathcal{V}_\mu \nu)(z)$ 
  is a so-called normalized Cauchy transform \cite{Cima2006a}.
  For objects like this, Aleksandrov's theorem allows one to compute the total detection probability \cite[Prop. 10.2.3]{Cima2006a}:
  \begin{equation}
    \TDP
    =
    \frac{1}{2\pi} \Int{0}{2\pi}{\lambda}
    \abs{
    ({\cal V}_\mu \nu)(e^{i\lambda})
    }^2
    =
    \frac{1}{2\pi}
    \Int{0}{2\pi}{\lambda}
    \mu(\lambda)
    \abs{\nu(\lambda)}^2 
    ,
  \label{eq:Aleksandrov}
  \end{equation}
  where we used Eq.~\eqref{eq:TDPHardy} for the first equality.  An explicit example of how Aleksandrov's theorem works in practice for a two-level system in given in App.~\ref{AppTwoLev}

  Since $\mu(\theta)$ is just a sum of delta functions at the quasienergies $\lambda_l$,
  and $\nu(\lambda_l)$ is known, the required integral is easily determined:
  \begin{align}
    \TDP
    = & \nonumber 
    \frac{1}{2\pi}
    \Int{0}{2\pi}{\lambda}
    \mu(\lambda)
    \abs{\nu(\lambda)}^2 
    \\ = & \nonumber 
    \sideset{}{'} \sum_l \ev*{\hat{P}_l}{\PsiDet}
    \abs{
      e^{-i\lambda_l} \frac{
        \mel*{\PsiDet}{\hat{P}_l}{\PsiIn}
      }{
        \ev*{\hat{P}_l}{\PsiDet}
      }
    }^2
    \\ = &
    \sideset{}{'} \sum_l 
    \frac{
      \abs{\sSum{m=1}{g_l} \ip{\PsiDet}{E_{l,m}} \ip{E_{l,m}}{\PsiIn}}^2
    }{
      \sSum{m=1}{g_l} \ip{\PsiDet}{E_{l,m}} \ip{E_{l,m}}{\PsiDet}
    }
    .
  \label{eq:Stupid}
  \end{align}
  We have thus obtained Eq.~\eqref{eq:Kessler}. 
  
  It is important to note that Aleksandrov's theorem in this form only applies to operators with a point spectrum.  
  If the spectrum has a continuous component, then Aleksandrov's theorem turns into an inequality \cite[Prop. 10.2.3]{Cima2006a}, and we have
  \begin{equation} \TDP
   \le    
   \frac{1}{2\pi} \Int{0}{2\pi}{\lambda} \mu(\lambda) \abs{\nu(\lambda)}^2
   .
   \label{eq:AleksandrovInEq}
  \end{equation}  
  Ref.~\cite{Thiel2018b} discusses $\mu(\lambda)$ in depth for systems with a continuous spectrum.
  (Note that this reference defines $\nu(\lambda)$ in a slightly different way.)
  This proof does not make any explicit reference to dark or bright states (except for the exclusion of dark levels) and so is not physically transparent. 
  We therefore present an alternate proof in the next section.
  
\section{A partition of the Hilbert space}
\label{sec:Dichotomy}
\subsection{Dark States}
  In this section we discuss the partition of the Hilbert space into a bright and dark part.
  In Sec.~\ref{sec:Strobo}, we defined dark states as those with $F_n = 0$, $S_n=1$ for all $n$. 
  In particular, we focus on stationary dark states, which are invariant under unitary evolution as well as under the detection attempts.
  During the course of the detection protocol only their phase is affected.
  Hence they remain dark for all times.
  In view of the diagonalization \eqref{eq:TEODiag}, there are two ways dark states can arise.

  \paragraph{Completely dark quasienergy levels} 
    Consider an quasienergy level $E_l$ that has no overlap in the detection state.
    That means none of the level's eigenstates  overlaps with $\ket{\PsiDet}$, i.e. $\hat{P}_l\ket{\PsiDet} = 0$ or $\ip*{E_{l,m}}{\PsiDet} = 0$ for all $m=1,\hdots,g_l$.
    Alternatively, one can write  $\Detect\ket{E_{l,m}} = 0$.
    We also denote these states as $\ket{\delta_{l,m}} = \ket{E_{l,m}}$.  If we take one of these as an initial state,
    we have: $\Surf\ket{\delta_{l,m}} = (\Id - \Detect) \TEO(\tau) \ket{\delta_{l,m}} = (\Id - \Detect) e^{-i\tau E_l/\hbar}\ket{\delta_{l,m}} = e^{-i\tau E_l / \hbar} \ket{\delta_{l,m}}$.
   Thus, $\ket{\delta_{l,m}}$ is an eigenstate of the survival operator with an eigenvalue on the unit circle,
    and so $S_n(\delta_{l,m}) = \norm*{\Surf^n\ket{\delta_{l,m}}}^2 = 1$. Thus, all the $\ket{\delta_{l,m}}$ are dark states.
    We call quasienergy levels that do not appear in a decomposition of the detection state {\em completely dark quasienergy levels}.
    All $g_l$ associated eigenstates are dark.

  \paragraph{Degenerate energy levels}
    Consider now an quasienergy level $E_l$ that does overlap with the detection state, but which is degenerate, such that $g_l > 1$.  Construct the projection of the detection state on this sector,
    \begin{equation}
    \ket{\beta_l} \equiv  \frac{\hat{P}_l\ket{\PsiDet}}{\sqrt{\smash[b]{\ev*{\hat{P}_l}{\PsiDet}}}}
    \label{eq:BrightEigen}
    \end{equation}
    It will turn out that $\ket{\beta_l}$ is a bright state.
    All states within this sector which are orthogonal to $\ket{\beta_l}$ are orthogonal to the detector, since if $\ket{\delta_l}$ is such a state,
    \begin{equation}
    \bra{\PsiDet}\ket{\delta_l} =   \bra{\PsiDet}\Detect\ket{\delta_l} \propto \bra{\beta_l}\ket{\delta_l} = 0
    \end{equation}
    and so are  dark.
    These states constitute a $g_l - 1$ dimension subspace of this quasienergy sector, any state lying within which is dark.
    It can be convenient to have an explicit basis for this subspace, which can be generated by a determinantal formula similar to that used  in the Gram-Schmidt procedure.
    We define
    \begin{equation}
      \ket{\delta_{l,j}}
      =
      N_j
      \mqty|
        \ket{E_1} & \ket{E_2} & \cdots & \ket{E_{j+1}} \\
        \ip{\PsiDet}{E_1} & \ip{\PsiDet}{E_2} & \cdots & \ip{\PsiDet}{E_{j+1}} \\
        \ip{\delta_1}{E_1} & \ip{\delta_1}{E_2} & \cdots & \ip{\delta_1}{E_{j+1}} \\
        \vdots & \vdots & \ddots & \vdots \\
        \ip{\delta_{j-1}}{E_1} & \ip{\delta_{j-1}}{E_2} & \cdots & \ip{\delta_{j-1}}{E_{j+1}}\\
      |
    \label{eq:StatDarkDet}
    \end{equation}
    where $N_j$ is a normalization factor and the subscript $l$ was omitted on the right-hand side.
    This determinant has the obvious properties that 
    (i) it yields an eigenstate of $\TEO(\tau)$;
    (ii) it is orthogonal to $\ket{\PsiDet}$, since $\ip*{\PsiDet}{\delta_{l,j}}$ is a determinant with 
    two identical rows; and (iii) that it is orthogonal to all $\ket{\delta_{l,m}}$ with $m<j$.
    For the same reason, 
    $\ip*{\delta_{l,m}}{\delta_{l,j}}$ is a determinant with two identical rows.

    This formula gives 
    \begin{equation}
    \ket{\delta_{l,1}}=N_1 \left(\ip*{\PsiDet}{E_{l,2}} \ket{E_{l,1}} - \ip*{\PsiDet}{E_{l,1}} \ket{E_{l,2}}\right)
    \label{eq:StationaryDark1}
    \end{equation}
    and then  recursively computes $\ket{\delta_{l,m}}$ from all previous $\ket{\delta_{l,j}}$.
    The result is a set of $g_l-1$ normalized and mutually orthogonal stationary dark states.
    As $\ket{\delta_{l,m}}$ is constructed only from the eigenstates $\ket{E_{l,1}}, \ket{E_{l,2}}, \hdots, \ket{E_{l,m+1}}$, 
    we find that many of the $\ip{\delta_{l,m}}{E_{l,j}}$ in Eq.~\eqref{eq:StatDarkDet} vanish, namely all those with $j > m+1$.
    Consequently the matrix in Eq.~\eqref{eq:StatDarkDet} is lower triangular except for the first rows.
    This enables us to compute the determinant explicitly:
    \begin{equation}
      \ket{\delta_{l,m}} 
      = 
      \frac{
        \sSum{j=1}{m} \qty[ 
          \abs{\alpha_{l,j}}^2 \ket{E_{l,m+1}} 
          - \alpha_{l,m+1}^*\alpha_{l,j} \ket{E_{l,j}}
        ]
      }{
        \sqrt{
          \sSum{j=1}{m} \sSum{j'=1}{m+1} 
          \abs{\alpha_{l,j}}^2
          \abs{\alpha_{l,j'}}^2
        }
      }
      .
    \label{eq:StationaryDark}
    \end{equation}
    Here we abbreviated $\alpha_{l,m} = \ip{E_{l,m}}{\PsiDet}$.
    Eq.~\eqref{eq:StationaryDark} contains Eq.~\eqref{eq:StationaryDark1} as a special case.
    
\subsection{Bright States}
We have seen above how dark states arise, and constructed a set of stationary dark states, which are not only energy eigenstates, but (unit modulus eigenvalue) eigenstates of the survival operator. We have also seen that
    each degenerate level that was not completely dark yielded an energy eigenstate  $\ket{\beta_l}$ which we claimed was not only not dark, but is bright. Similarly, every non-degenerate level that is not completely dark also turns out to be bright, so that $\ket{E_l}=\ket{\beta_l}$. These bright states, while energy eigenstates, are not in general eigenstates of the survival operator.  Together, the $\ket{\beta_l}$ states, one arising from each not-totally-dark level, span the orthogonal complement of the dark space. 
    
    We now demonstrate that they are indeed bright.   It was already mentioned that the stationary dark state $\ket{\delta_{l,m}}$ is a (right-)eigenstate 
    of $\Surf$ with eigenvalue $e^{-i \tau E_l / \hbar}$, lying on the unit circle. It is easy to see that these are the only
    eigenvalues on the unit circle, since $\hat{D}$ must annihilate it. 
    Since $\Surf$ is the product of a projector with a unitary matrix, it can have no eigenvalues outside the unit disk,
    and so all other eigenvalues $\zeta$ must lie inside the unit disk, i.e. $\abs{\zeta} < 1$. 
    
    In fact, these eigenvalues are directly related to the poles of the generating function $\FDA(z)$ discussed above.
     We may now use the definition $\Surf := (\Id - \Detect) \TEO(\tau)$ of the survival operator and rewrite its characteristic polynomial
  as $\determinant[ \zeta \Id - \TEO(\tau) + \dyad{\PsiDet} \TEO(\tau) ]$.
  An application of the matrix determinant lemma yields:
  \begin{align}
    \determinant[ \zeta\Id - \Surf]
    = &
    \determinant[ \zeta\Id - \TEO(\tau) ] \ev*{[\zeta\Id - \TEO(\tau)]^{-1}}{\PsiDet}
    .
    \label{eq:Determ}
  \end{align}
  The last term can be identified with the denominator of $\FDA(z)$ of Eq.~\eqref{eq:DefGenFunc}.
  Inverting the equation yields:
  \begin{equation}
    \ev*{\tfrac{\Id}{\Id - \frac{1}{\zeta}\TEO(\tau)}}{\PsiDet}
    =
    \zeta
    \frac{\determinant[\zeta \Id - \Surf]}{\determinant[ \zeta \Id - \TEO(\tau) ]}
    .
  \label{eq:FDADenom}
  \end{equation}
  This relation shows that $z_p$ is a (finite) pole of $\FDA(z)$ if and only if $\zeta_l = 1/z_p$ is an eigenvalue of $\Surf$, but not of $\TEO(\tau)$.
  These are exactly the eigenvalues of $\Surf$ that lie inside the unit disk.
  $\FDA(z)$ is by construction analytic in the unit disk.
  Therefore all of its poles must lie outside the unit disk and have $\abs{z_p} > 1$.
  Consequently all of these non-trivial eigenvalues of $\Surf$ -- which are the poles' reciprocals -- must lie inside the unit circle, $\abs{\zeta_l} < 1$.
  The corresponding eigenvectors belong to the complement of the dark space, i.e., the bright space. 
  These vectors span the bright space, and it is clear that any superposition of bright eigenstates 
  $\ket*{\widetilde{\beta}} = \sSum{l}{} b_l \ket{\beta_l}$ will yield an exponentially decaying survival probability.
  So $S_\infty(\widetilde{\beta}) = 0$ and the state is bright, as claimed. 
  The decay rate is determined by the eigenvalue $\zeta_{\text{max}}$ closest to the unit circle.  
  
  There are two subtleties in this argument. One must allow for the possibility that there are degenerate eigenvalues of $\Surf$, and the operator is not diagonalizable. Showing that $\ket*{\widetilde{\beta}}$ is bright in this case is a bit technical, and we present the argument in Appendix~\ref{app:Surf}. The other point to be raised is that if the system is infinite, the eigenvalues might approach arbitrarily closely to the unit circle, and there might not be an exponential decay of the survival probability.  This indeed happens, and in such systems, states in the complement to the dark space are not necessarily bright. This would be discussed in more detail in Sec.~\ref{sec:Line}.
  
    The just described transition from $\{ \ket{E_{l,m}} \}_{m=1}^{g_l}$ to $\ket{\beta_l}$ and $\{ \ket{\delta_{l,m}} \}_{m=1}^{g_l-1}$
    is a change from one orthonormal basis to another.
        Still, all involved states are eigenstates, and are thus invariant under $\TEO(\tau)$.
    The special feature of the new representation is that each individual stationary dark state and all bright eigenstates together are additionally invariant under the detection process.
    In the language of Refs.~\cite{Facchi2003a, Caruso2009a}, $\Hilbert_B$ and $\Hilbert_D$ are so-called invariant subspaces.
    The action of the survival operator $\Surf$ may change a particular superposition of bright eigenstates 
    $\ket*{\widetilde{\beta}} = \sSum{l}{} b_l \ket{\beta_l}$ into some other superposition of $\ket{\beta_l}$, 
    but it can never generate a dark state.

  \section{$\TDP$ calculated from the Bright Space\label{sec:TDP2}}
    We see from the above discussion that, as noted by Krovi and Brun~\cite{Krovi2006a},  the limit $\lim_{n\to\infty}S_n(\PsiIn)$ of an arbitrary initial state's survival probability is equal to its overlap with the dark space:
    $\norm*{\hat{P}_{\Hilbert_D}\ket{\PsiIn}}^2$, at least for a finite system.
    Complementarily, the total detection probability must be equal to the overlap with the bright space.
    
    The projector onto the bright space has the form: $\hat{P}_{\Hilbert_B} = \sSum{l}{'} \dyad{\beta_l}$, 
    where the sum excludes all completely dark energy levels, so that they do not possess a bright state, and $\ket{\beta_l}$ are the bright eigenstates of Eq.~\eqref{eq:BrightEigen}.
    Identifying $\TDP(\PsiIn) = \ev*{\hat{P}_{\Hilbert_B}}{\PsiIn}$, we find:
    \begin{equation}
      \TDP(\PsiIn)
      =
      \sideset{}{'} \sum_l
      \frac{\abs*{\mel*{\PsiDet}{\hat{P}_l}{\PsiIn}}^2}{\ev*{\hat{P}_l}{\PsiDet}}
      .
    \label{eq:PreKessler}
    \end{equation}
    Plugging the definition of the eigenspace projectors $\hat{P}_l$ that is found in Eq.~\eqref{eq:TEODiag} into Eq.~\eqref{eq:PreKessler}, yields Eq.~\eqref{eq:Kessler}.
    Note that the sum only runs over those quasienergy levels $E_l$ which are not completely dark,    so that the denominator in Eq.~\eqref{eq:Kessler} can never vanish.

    We can now make the following observations:

    (a) When the initial state coincides with the detection state, we get $\TDP = 1$ in agreement with Ref.~\cite{Gruenbaum2013a}.
    This means that the detection state $\ket{\PsiDet}$ itself is always bright.

    (b) The total detection probability does not depend on the sampling rate.
    $\TDP$ is $\tau$-independent.
    This is true except for the resonant detection periods $\tau_c$, defined in Eq.~\eqref{eq:DefResonantTau}.
    At these points the number of bright states suddenly changes and $\TDP$ will exhibit a discontinuous drop.

    (c) In systems with only non-degenerate quasienergy levels that all overlap with the detection state,
    i.e. $\ip{E_l}{\PsiDet} \ne 0$, any initial state is detected with probability one.
    In this case there are no dark states, all eigenstates are bright.
    Such a behavior is classical, in the sense that a classical ergodic
    random walk also finds its target with probability one.
    It implies that some disorder in the system, which removes all the degeneracy,
    may increase the detection probability provided
    that none of the eigenstates is orthogonal to the detection state.
    In systems with degenerate quasienergy levels, one may still be able to find initial states which are 
    detected with probability one, but this will not be true generically for {\em all} initial states.

     \section{Examples\label{sec:Examples}}

  Let us now consider some examples that demonstrate the relation of $\TDP$ to the dark and bright states.
  The examples all consist of graphs of nodes connected by links, which represent hopping between the connected nodes, and we consider only initial and detection states which are localized on individual nodes. In all cases but the ring with a magnetic field, we take the hopping strengths to be uniform, with value $\gamma$, and zero on-site energies. We also assume that $\tau$ avoids the resonant condition \eqref{eq:DefResonantTau}.
  
   \subsection{A ring}
    Let us now consider a ring with $L$ (even) sides.
    Its Hamiltonian takes the form:
    \begin{equation}
      \Ham
      =
      - \gamma \Sum{r=1}{L} [ \dyad{r}{r+1} + \dyad{r}{r-1} ]
      .
    \label{eq:RingHam}
    \end{equation}
    We employ periodic boundary conditions and identify $\ket{r+L} = \ket{r}$.
   
    The free wave states diagonalize the Hamiltonian, such that the energy levels are given by:
    \begin{equation}
      E_l
      =
      - 2 \gamma \cos\frac{2\pi l}{L}
      ,
    \label{eq:RingE}
    \end{equation}
    with $l=0,1,\hdots,L/2$.
    Except for $E_0$ and $E_{L/2}$, which are non-degenerate, all energy levels possess two eigenstates:
    \begin{align}
      \left\{ \begin{aligned}
        \ket{E_0} = \Sum{r=1}{L} \frac{\ket{r}}{\sqrt{L}}
        \qc
        \ket{E_{L/2}} = \Sum{r=1}{L} \frac{(-1)^r \ket{r}}{\sqrt{L}}
        \\
        \ket{E_{l,1}} = \Sum{r=1}{L} \frac{e^{i \frac{2\pi l r }{L}} \ket{r}}{\sqrt{L}}
        \qc
        \ket{E_{l,2}} = \Sum{r=1}{L} \frac{e^{-i \frac{2\pi l r }{L}} \ket{r}}{\sqrt{L}}
      \end{aligned} \right.
      .
    \label{eq:RingStatStates}
    \end{align}
    Picking the localized states $\ket{\RIn}$ and $\ket{\RDet} = \ket{L}$ as initial and detection state,
    we find, using Eq.~\eqref{eq:Kessler}
    \begin{equation}
      \TDP(\RIn)
      =
      \frac{2}{L}
      +
      \frac{2}{L}
      \Sum{l=1}{\frac{L}{2}-1}
      \cos^2\frac{2\pi l \RIn}{L}
      =
      \left\{ \begin{aligned}
        1
        \qc & \RIn = \tfrac{L}{2},L
        \\
        \frac{1}{2}
        \qc & \text{else}
      \end{aligned} \right.
      .
    \label{eq:TDPRing}
    \end{equation}
    For almost all sites, we find $\TDP = 1/2$ except when $\RIn$ and $\RDet$ coincide or are on opposing sides of the ring.
    This is in accordance to our results from Ref.~\cite{Friedman2017b}.
    The same result appears for rings of odd sizes, albeit it is not possible to place initial and detection site on opposite sites of the ring.

    The energy basis bright states are given by the cosine waves, whereas the stationary dark states are given by sine waves:
    \begin{equation}
      \ket{\beta_l}
      =
      N_l
      \Sum{r=1}{L} \cos\frac{2\pi l r}{L} \ket{r}
      \qc
      \ket{\delta_l}
      =
      \sqrt{\frac{2}{L}}
      \Sum{r=1}{L} \sin\frac{2\pi l r}{L} \ket{r}
      ,
    \label{eq:RingStates}
    \end{equation}
    where $N_l = \sqrt{1/L}$ for $l=0,L/2$ and $N_l = \sqrt{2/L}$ otherwise.
    $\ket{\delta_0}$ and $\ket{\delta_{L/2}}$ are not defined.  Thus, in this case, the deviation of $\TDP$ from unity is solely due to the degeneracies of the energy spectrum, which of course arise from the parity symmetry of the system.

  \subsection{A Ring with a magnetic field}
      \begin{figure}
        \includegraphics[width=0.9\columnwidth]{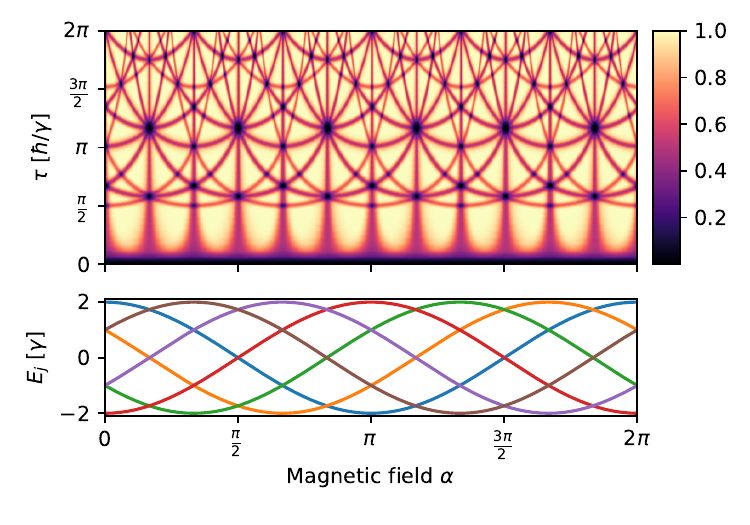}
        \caption{
          Total detection probability for a ring with magnetic field.
          Here $\TDP$ was approximated by $\TDP = 1 - S_{N}$, with $N=50$.
          Initial and detection state are: $\RDet = 0$ and $\RIn = 1$.
          Bottom: The energy levels of Eq.~\eqref{eq:MagRingEnergyLevels} as a function of $\alpha$.
          Degeneracy occurs when they cross, this occurs for special values of $\alpha$.
          Top: The total detection probability as a function of $\tau$ and $\alpha$.
          We usually find $\TDP = 1$, but a deficit occurs when there is a degeneracy in the quasienergy levels which appears as dark lines in the figure.
          This happens for certain values of $\alpha$ (vertical lines, coinciding with the intersection of two energy levels in the bottom) 
          and for special combinations of $\tau$ and $\alpha$.
          As $N$ is taken to infinity the dark lines will become infinitely thin.
          \label{fig:Dror}
        }
      \end{figure}
      In the ring, as we saw, the degeneracy of an energy level gives rise to dark states,
      which in turn lead to a deficit in the total detection probability.
      What happens if this degeneracy is lifted?
      To explore this, we  add to the ring model of Eq.~\eqref{eq:RingHam}  a magnetic field:
      \begin{equation}
        \Ham
        =
        - \gamma \Sum{r=1}{L} \left( e^{i\alpha} \dyad{r}{r+1} + e^{-i\alpha}\dyad{r}{r-1} \right)
        .
      \label{eq:MagRingHam}
      \end{equation}
      The magnetic field strength is proportional to $\alpha$ and its vector is normal to the plane in which the ring lies.
      This field splits up the two-fold degeneracy of the energy levels.
      The eigenstates are still the free wave states of Eq.~\eqref{eq:RingStatStates}, although the index $l$ now runs from zero to $L-1$.
      The new levels are 
      \begin{equation}
        E_l
        = 
        -2 \gamma \cos(\frac{2\pi l}{L} + \alpha)
        .
      \label{eq:MagRingEnergyLevels}
      \end{equation}
      Except for special values of $\alpha$, namely when $\alpha$ is an integer multiple of $2\pi/L$, these are all distinct.
      Similarly, a degenerate quasienergy level can only appear when condition \eqref{eq:DefResonantTau} for resonant $\tau$ is fulfilled.
      Since a localized detection state $\ket{\RDet}$ has overlap with all eigenstates and since all energy levels are non-degenerate, we find
      \begin{equation}
        \TDP(\RIn) = 1
      \label{eq:TDPRingMag}
      \end{equation}
      with the exception of special combinations of $\tau$ and $\alpha$.
      This is nicely demonstrated in Fig.~\ref{fig:Dror}, $\TDP$ as a function of $\tau$ and $\alpha$,
      as well as the energy levels as a function of $\alpha$ for a ring with $L=6$.
      We find that the total detection probability is almost everywhere unity except along some lines in $(\alpha,\tau)$-space which parametrize the resonant combinations at which degeneracy occurs. What is presented is $1-S_{50}$, rather than $1-S_\infty$. Thus, the lines have finite width. Nevertheless, the extreme sensitivity to the presence of any symmetry breaking is apparent.
      
\begin{figure}
  \begin{center}
  \includegraphics[width=0.9\columnwidth]{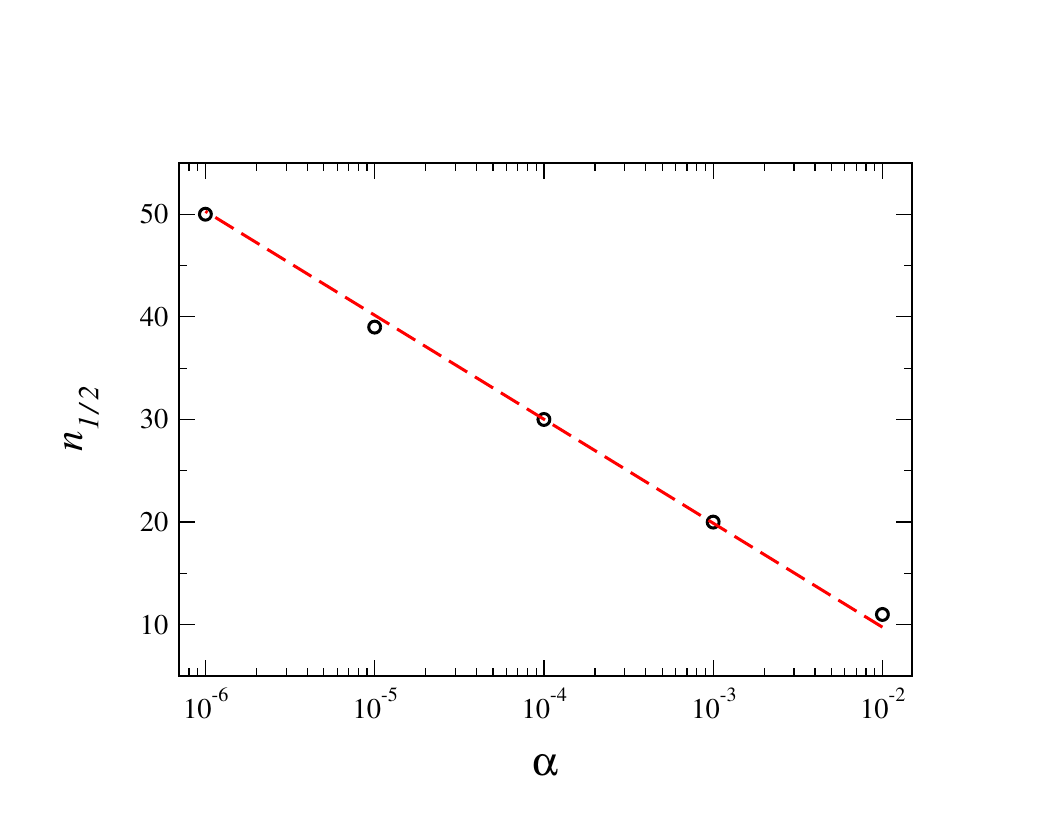}
  \end{center}
  \caption{$n_{1/2}$, the first $n$ for which $S_n(\alpha)<S_n(\alpha=0)=1/2$, as a function of $\alpha$, for a ring of size $L=6$, with $\RDet = 0$ and $\RIn = 1$, $\gamma\tau=\hbar$. A straight line is shown indicating the linear behavior in $\ln \alpha$ for small $\alpha$.}
  \label{fig:ringnc}
  \end{figure}

\begin{figure}
  \begin{center}
  \includegraphics[width=0.9\columnwidth]{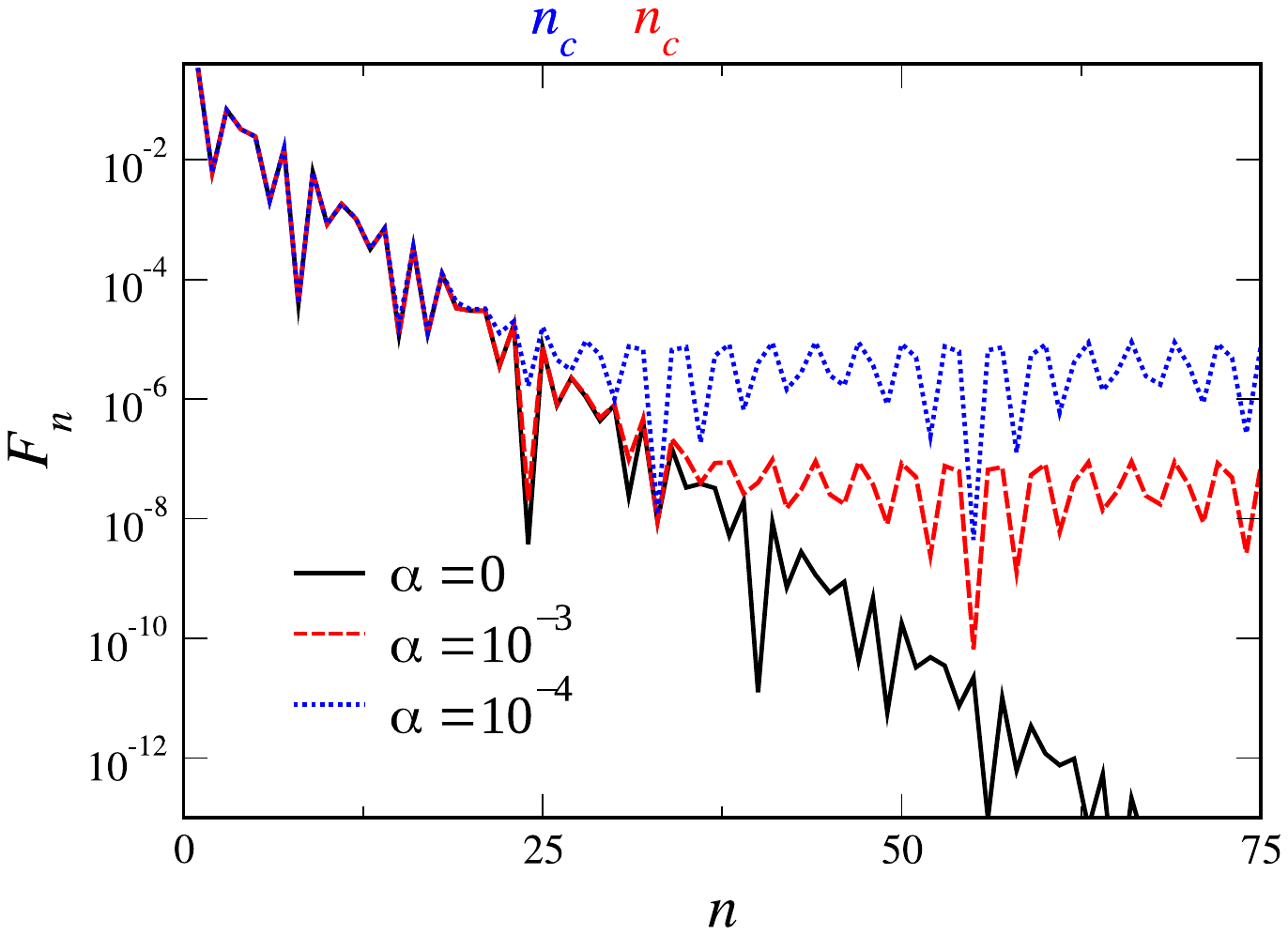}
  \end{center}
  \caption{The first detection probability $F_n$  for a $L=6$ ring with magnetic field parameter $\alpha=0$, $10^{-4}$ and $10^{-3}$. The behavior for small $n$ is indistinguishable in the three cases. Past some $n_c(\alpha)$, as marked on the figure, the finite $\alpha$ cases diverge from that of $\alpha=0$, decaying at a much slower rate.  Here, $\RDet = 0$ and $\RIn = 1$, $\gamma\tau=\hbar$.}
  \label{fig:RingFn}
  \end{figure}

        This can be quantified by considering the value of $n_{1/2}$ for which $S_n$ falls below its $\alpha=0$ limit of 1/2 (for the detection site not identical or diametrically opposite the initial site). This is plotted in Fig. \ref{fig:ringnc}, where even for tiny value of $\alpha=10^{-6}$, the presence of a magnetic field can be unequivocally detected after a series of runs of 50 measurement attempts.
      We see that $n_{1/2}$ grows only logarithmically with diminishing $\alpha$, so that detection of very small magnetic fields is in principle possible. This can be understood by looking at the behavior of $F_n$, as depicted in Fig. \ref{fig:RingFn}. We see that, for the small values of $\alpha$  depicted,  $F_n$ is essentially independent of $\alpha$ for not too large $n$. For small, finite $\alpha$ and large $n$, $S_n$ decays very slowly, as a result of  the presence of slow modes of $\Surf$. These modes, which for $\alpha=0$ were dark states with unit modulus eigenvalue, now  are  at a distance of order $\alpha^2$ from the unit circle, leading to a decay rate of order $\alpha^2$. These modes have to contribute a total of $1/2$ to $\TDP$. Given their slow ${\cal{O}}(\alpha^2)$ exponential decay, once this slow decay sets in, (at $n_c$, say), 
      $F_n$ must behave as $F_n \sim F_{n_c} \exp\left(-d \alpha^2 (n-n_c)\right)$, for some $\alpha$-independent $d$, with $F_{n_c} \sim \alpha^{-2}$, to give an $\alpha$-independent sum. Thus, in Fig. \ref{fig:ringnc} we see that $F_{n_c}$ for $\alpha=10^{-4}$ is roughly a factor of 100 smaller than for $\alpha^{-3}$.
   Since, for $n<n_c$, the decay rate of $F_n$ is set by the slowest non-dark state (with an $\alpha$-independent decay rate, $d_0$), we get that $F_{n_c} \sim \exp(-d_0 n)$, so that 
   \begin{equation}
   n_c \propto -\ln \alpha
   \end{equation}
Since, for $\alpha=0$, $\TDP=1/2$, $1-S_n$ will first exceed this value at an $n_{1/2}$ of order $n_c$.

  \subsection{Square with center node}
    We next consider a square with a single node in the middle connected to the all others.
    The states localized on the corners of the square are denoted by $\ket{1},\ket{2},\ket{3}$
    and $\ket{4}$, and the additional node in the center of the diagonals is state $\ket{0}$.
    The Hamiltonian in matrix form is 
    \begin{equation}
      \Ham = -\gamma  \mqty(
        0 & 1 & 1 & 1 & 1 \\
        1 & 0 & 1 & 0 & 1 \\
        1 & 1 & 0 & 1 & 0 \\
        1 & 0 & 1 & 0 & 1 \\
        1 & 1 & 0 & 1 & 0 
      ).
    \end{equation}
    The energy levels are $E_1=-(1 + \sqrt{5})\gamma$, $E_2 = (\sqrt{5}-1)\gamma$, $E_3 = 2\gamma$, and $E_4 = 0$, 
    and $E_4$ is two-fold degenerate.
    To find $\TDP$ we obtain the eigenvectors:
    \begin{align}
      \left\{ \begin{aligned}
        \ket{E_1}
        = &
        ( \sqrt{5} - 1, 1, 1, 1)^T / \sqrt{10 - 2\sqrt{5}}
        , \\
        \ket{E_2}
        = &
        (-1-\sqrt{5}, 1, 1, 1, 1)^T / \sqrt{10 + 2\sqrt{5}}
        , \\
        \ket{E_3}
        = &
        (0, -1 , 1 , -1 , 1 )^T / 2
        ,
        \\
        \ket{E_{4,1}}
        = &
        (0, 1, -1, -1 , 1)^T / 2
        , \\ 
        \ket{E_{4,2}}
        = &
        (0, 1, 1, -1, -1)^T / 2
      \end{aligned} \right.
      .
    \label{eq:SquareStates}
    \end{align}
    Notice that, as opposed to the ring, where there were no completely dark levels, here there are two energy levels, $E_3$ and $E_4$, which are    completely dark with respect to a measurement in the center. These sectors are dark because these states are not invariant under rotation by $\pi/2$ (where $\ket{E_3}$ transforms to $-\ket{E_3}$, and $\ket{E_{4,1}}$
    transforms to $-\ket{E_{4,2}}$), whereas the center is, and so there can be no overlap with the center. As per our general discussion, this gives rise to a  deviation of $\TDP$ from unity
    when measuring at exactly this node, i.e. $\ket{\RDet} = \ket{0}$.
    The bright eigenstates are given by the projections of the detection state into the energy subspaces,
    hence $\ket{\beta_1} = \ket{E_1}$ and $\ket{\beta_2} = \ket{E_2}$.
    As mentioned the energy levels $E_3$ and $E_4$ are completely dark and will be excluded in 
    the sum of Eq.~\eqref{eq:Kessler} that yields:
    \begin{equation}
      \TDP(\RIn)
      =
      \left\{ \begin{aligned}
          1 \qc & \RIn = 0 \\
          \tfrac{1}{4} \qc & \RIn \ne 0
      \end{aligned} \right.
      .
    \label{eq:TDPSquare}
    \end{equation}
   
    For other detection states, say
   $\RDet = 1$, as other initial states are the same up to a rotation,
    there are no dark energy levels,
    but $E_4$ possesses one dark state due to degeneracy: $\ket{\delta_4} := (\ket{E_{4,1}} - \ket{E_{4,2}})/\sqrt{2}$.
    All energy levels participate in Eq.~\eqref{eq:Kessler}:
    \begin{equation}
      \TDP(\RIn)
      =
      \left\{ \begin{aligned}
        1 \qc & \RIn = 0, 1, 3 \\
        \tfrac{1}{2} \qc & \RIn = 2,4
      \end{aligned} \right.
      .
    \label{eq:TDPSquareOuter}
    \end{equation}

\subsection{Additional examples}
\begin{figure*}
    \centering
    \includegraphics[width=0.99\textwidth]{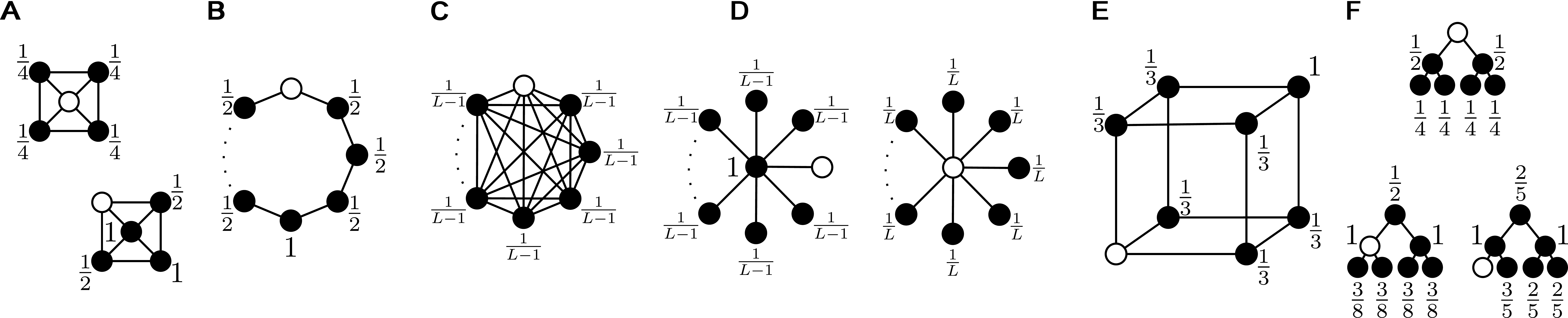}
    \caption{
      Total detection probability for graphs.
      A quantum particle is put on a graph, whose node's describe the particle's possible position states.
      The links describe allowed transitions each with equal strength.
      The particle starts localized on some node of the graph (full circles) and we attempt to 
      detect it on the node with the open circle.
      The numbers denote $\TDP$ for this initial state.
      (A) Square with a center node,
      (B) ring of $L$ sites,
      (C) complete graph with $L$ sites,
      (D) star graph with $L$ sites in the periphery
      (E) hypercube (here of dimension $d=3$
      (F) binary tree with $2$ generations.
      \label{fig:ExampleGraphs2}
    }
  \end{figure*}

We have worked out a number of additional examples where $\TDP<1$ and the associated dark states for other simple geometries of graphs. These include a complete graph, a star graph, a hypercube and a tree graph. The results are presented in  appendix C, and summarized together with the previous examples in Fig. \ref{fig:ExampleGraphs2}.
The detection node is represented by an open circle,
  All other (filled) circles are possible initial states and the numbers next to them represent the corresponding total detection probability.
 It is interesting to note that in all these simple cases, $\TDP$ is rational, and except for the tree, is simply 1 over an integer. 
The detection probability for some  Larger systems is discussed in Ref.~\cite{Thiel2020a}.

We note that
  after infinitely many unsuccessful detection attempts only the dark component of the initial wave function survive.
  The dark space is invariant under unitary evolution and strong detection \cite{Facchi2003a, Schaefer2014a}.
  Similar behavior can be found in the long-time behavior of general open quantum dynamics \cite{Novotny2012a}.
  An interesting question would be which of the here described features will carry over when one
  employs weak measurements or open quantum dynamics \cite{Svensson2013a, Tamir2013a, Gurvitz2017a, Chan2019a}.

  \section{The Infinite Line\label{sec:Line}}
    We have seen above that finite systems are very different than infinite systems, at least as far as the properties of $\TDP$ are concerned.
    Gr\"unbaum, et al.~\cite{Gruenbaum2013a} showed that, when the detection and initial states coincide, $\TDP$ is  unity for every finite system, for an infinite system with a band of continuous energies, such as the infinite line,  it is always less than unity~\cite{Friedman2017b}.
    We wish here to amplify on this point by comparing the finite ring of size $L$ with its infinite counterpart.
    Above we exhibited the dark states for the finite ring.
    It is clear that the dark states remain dark even in the infinite-$L$ limit.
    However, the bright states of the finite ring, while remaining orthogonal to every dark state, are not detected with unit probability in the infinite-$L$ limit.
    Instead, they are ``dim'', such that $0 < \TDP < 1$ for these states.
    Furthermore, $\TDP$ for these states depends essentially on $\tau$.  
 
\begin{figure}
    \centering
    \includegraphics[width=0.49\textwidth]{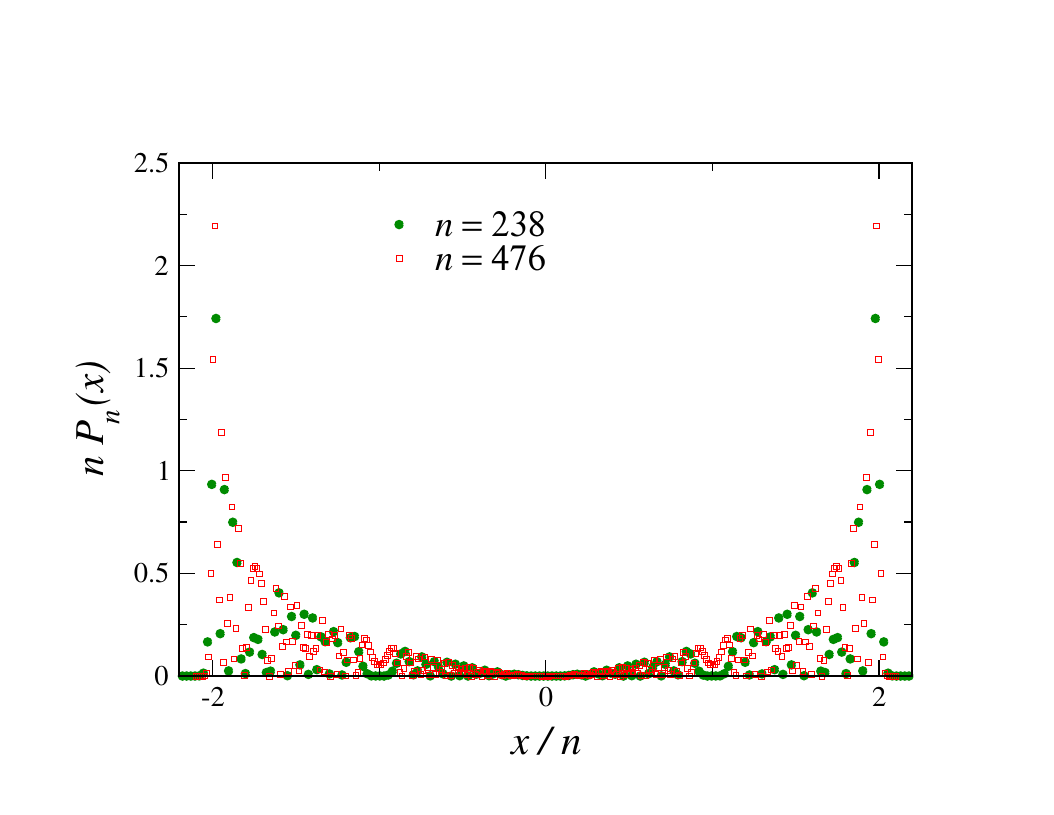}
    \caption{
      The surviving (i.e., as yet undetected) particle probability density $P_n(x)$ after $n=238$ and $576$ measurements, in a ring of length $L=2000$ and a detection interval of $\gamma \tau/\hbar = 1$. For clarity, only every 5th point is plotted.
            \label{fig:spread}
    }
  \end{figure}

 To get an insight into this phenomenon, we first consider the space-time picture of the undetected probability density (i.e., the position distribution at time $n\tau$, normalized to $S_n$).  We start with the particle at $x=L/2$ on a ring of large length $L$, measuring at the same point. As time progresses, the density spreads out ballistically from the initial location in both directions.  
 As long as not enough time has passed to allow the two ``wings" of the distribution to meet at the opposite end of the ring ($x=0=L$), a time of order $L$, the density has a overall amplitude proportional to $1/t$, so that, given the extension over a distance proportional to $t$,  the total undetected density is of order unity. It should be noted that this is the same exact scaling behavior of the  measurement-free density. This situation is presented in Fig. \ref{fig:spread}, where the scaled density is presented after $238$ and $476$ steps. The total undetected probability, $S_n$, is $0.6006$ at both times, the difference being only of order $4\cdot 10^{-6}$, while the average distance to the origin of the surviving particle has doubled from $386$ to $770$. Thus, in the limit of large $L$,  the total undetected probability remains of order unity, spreading out to infinity, for all times less than the huge system traversal time of order $L$.
 
 An alternative picture arises from considering the stationary bright states which are the eigenvectors of the survival operator with eigenvalues lying inside the unit circle. For large finite $L$, these eigenvalues span the range in magnitude from 0 to very near unity.  For times $N\tau$ shorter than the time to propagate across the system, (i.e., for the wings of the distribution to meet, as described above) the relatively quickly decaying stationary bright states with eigenvalues with small magnitude contribute to $\sum_{n\le N}F_n =1-S_N$, while the states with eigenvalues near the unit circle do not.
 These slow states, then, are essentially dark over this time-scale. We find that for $L=1000$, there are 267 fast eigenvalues (out of 501) further than a distance .002 from the unit circle, and for $L=2000$, there are 531 fast eigenvalues (out of 1001) further than a distance .001 from the unit circle (for $\gamma\tau=\hbar$); i.e., twice as many.  Examining the eigenvectors, they extend over the entire system, so that an initial state localized near the origin will have a squared overlap with each bright mode proportional to $1/L$. Thus, the number of effectively dark states, those that do not contribute in time $N\tau$ to $\TDP$, scales with $L$. Thus, the undetected probability is of order 1, since it arises from order $L$ modes times a number of order $1/L$ overlap for each mode.

\section{From Symmetry to Detection Probability}

 As mentioned in the introduction, the  symmetry of the underlying graph controls
the deviations of the detection probability from its classical counterpart which is unity 
    \cite{Krovi2006a}.
This is also evident in the  examples in 
      Fig. \ref{fig:ExampleGraphs2}
which clearly show  that symmetry plays an important role. 
Take  example B:  the ring. 
Then the probability to detect the particle once starting at one of the nearest 
neighbors of the detected site is $1/2$. 
This, as we claim below,  is related to the fact that we have two nearest neighbors on the ring structure. On the other hand,
 starting on the site opposing the detected site
yields $P_{{\rm det}}=1$. 
We claim that this is due to the absence of another equivalent initial
state in the system. 
Similarly for example E: the cube. Here, starting on one 
of the three nearest neighbors to the detector we find $P_{{\rm det}}=1/3$.
In the remaining   part of the paper we wish to further
 explore this relation between symmetry and $P_{{\rm det}}$.

  In this section we assume that the initial state is localized:
  \begin{equation}
    \ket{\PsiIn} = \ket{\RIn}
    .
  \label{eq:}
  \end{equation}
 We already discussed the first detection amplitude  
given by 
\cite{Dhar2015a, Dhar2015b, Friedman2017b}:
  \begin{equation}
    \FDA_n(\PsiIn)
    =
    \mel*{\PsiDet}{\TEO(\tau)[(\Id-\Detect) \TEO(\tau)]^{n-1}}{\PsiIn}
    .
  \label{eq:DefFDA}
  \end{equation}
  Reading the equation right-to-left, we see that the initial state
  is subject to $n-1$ double steps of unitary evolution and unsuccessful detection,
  until finally, after the $n$-th evolution step, detection is successful.
As mentioned the focus of this paper is the total detection probability
$    \TDP(\PsiIn)
    =
    \sSum{n=1}{\infty} F_n(\PsiIn)
    =
    \sSum{n=1}{\infty} \abs{\FDA_n(\PsiIn)}^2$
    .

  From Eq.~\eqref{eq:DefFDA} we see that the detection amplitude $\FDA_n$ is linear with respect
  to the initial state, and thus obeys a superposition principle.
  This allows us to obtain an upper bound for $\TDP(\RIn)$ without the need of detailed calculations.
  To see this, consider localized initial and detection states $\ket{\RIn}$ and $\ket{\RDet}$.
  Now we introduce an auxiliary initial state which is a linear combination
  of any two different localized states
  \begin{equation}
    \ket{\AUS_\alpha}
    =
    \frac{1}{\sqrt{2}}
    ( \ket{\RIn} + e^{i\alpha} \ket{r'} )
    ,
  \label{eq:SuperPosInitial}
  \end{equation}
  where $\ip{\RIn}{r'} = 0$ and $\alpha$ is an arbitrary relative phase.
  As the first detection amplitudes are linear in the initial state, see Eq.~\eqref{eq:DefFDA}, we find
  \begin{equation}
    \FDA_n(\AUS_\alpha)
    =
    \frac{1}{\sqrt{2}} \qty[
      \FDA_n(\RIn)
      + e^{i\alpha} \FDA_n(r')
    ]
    .
  \label{eqRE10}
  \end{equation}

  We can use this to find a very useful upper bound on the detection probability,
  provided there is a symmetry relation between $\RIn$ and $r'$, namely, when $\ket{\RIn}$ and $\ket{r'}$
  are physically equivalent. Mathematically, two orthogonal initial states 
$\ket{\PsiIn}$ and
  $\ket{\PsiIn'}$ which yield identical transition amplitudes to the detection  
\begin{equation}
    \mel*{\RDet}{\TEO(t)}{\PsiIn}
    =
    \mel*{\RDet}{\TEO(t)}{\PsiIn'}
    \ne 0
    .
  \label{eq:DefPhysEq}
  \end{equation}
are called physically  equivalent to each other. As usual, the amplitudes on the left and right hand side of this equation may have irrelevant phase factors
$e^{i \lambda}$. 
More importantly, such states only appear in systems with a certain degree
of symmetry. 
 From these two states we can construct the superposition state $(\ket{\PsiIn} - \ket{\PsiIn'})/\sqrt{2}$,
  which  must be a dark state.
  Hence, any pair of physically equivalent states yields a dark state.
  This reveals dark states as an interference phenomenon 
    \cite{Krovi2006a}.
  Certain initial states result in permanent destructive interference and thus vanishing probability amplitude in the detection state.

   Let us consider the example in the ring system of Fig.~\ref{fig:PD}, the two sites left and right of the detection node
  are equivalent due to reflection invariance.
  Clearly, when $\ket{\RIn}$ and $\ket{r'}$ are physically equivalent then also $\FDA_n(\RIn) = \FDA_n(r')$.
  Under such circumstances, for the superposition $\ket{\AUS_\alpha}$ in Eq.~\eqref{eq:SuperPosInitial}, we have:
  \begin{equation}
    \FDA_n(\AUS_\alpha)
    =
    \frac{1 + e^{i\alpha}}{\sqrt{2}}
    \FDA_n(\RIn)
  \label{eqRE11}
  \end{equation}
  and so
  \begin{equation}
    F_n(\AUS_\alpha)
    =
    \abs{ \FDA_n(\AUS_\alpha) }^2
    =
    ( 1 + \cos\alpha ) F_n(\RIn)
  .
  \label{eqRE11}
  \end{equation}
  The relative phase $\alpha$ affects the first detection statistics.
  The aforementioned  choice $\alpha=\pi$ obviously yields a dark state as can be seen.
  A useful bound on $\TDP$ can be found by summing over all $n$ and using $\TDP(\AUS_\alpha) = \sSum{n=1}{\infty} F_n(\AUS_\alpha) \le 1$:
  \begin{equation}
    1 \ge
    \TDP(\AUS_\alpha)
    =
    (1 + \cos\alpha) \TDP(\RIn)
    .
  \label{eqRE12}
  \end{equation}
  Choosing $\alpha=0$, we obtain for the originally considered transition from $\RIn$ to $\RDet$:
  \begin{equation}
    \TDP(\RIn)
    \le
    \frac{1}{2}
    .
  \label{eqRE13}
  \end{equation}
  Thus, if $\RIn$ has a physically equivalent partner, the total detection probability cannot be unity,
  unlike for the classical random walk.

  \begin{figure}
    \centering
    \includegraphics[width=0.99\columnwidth]{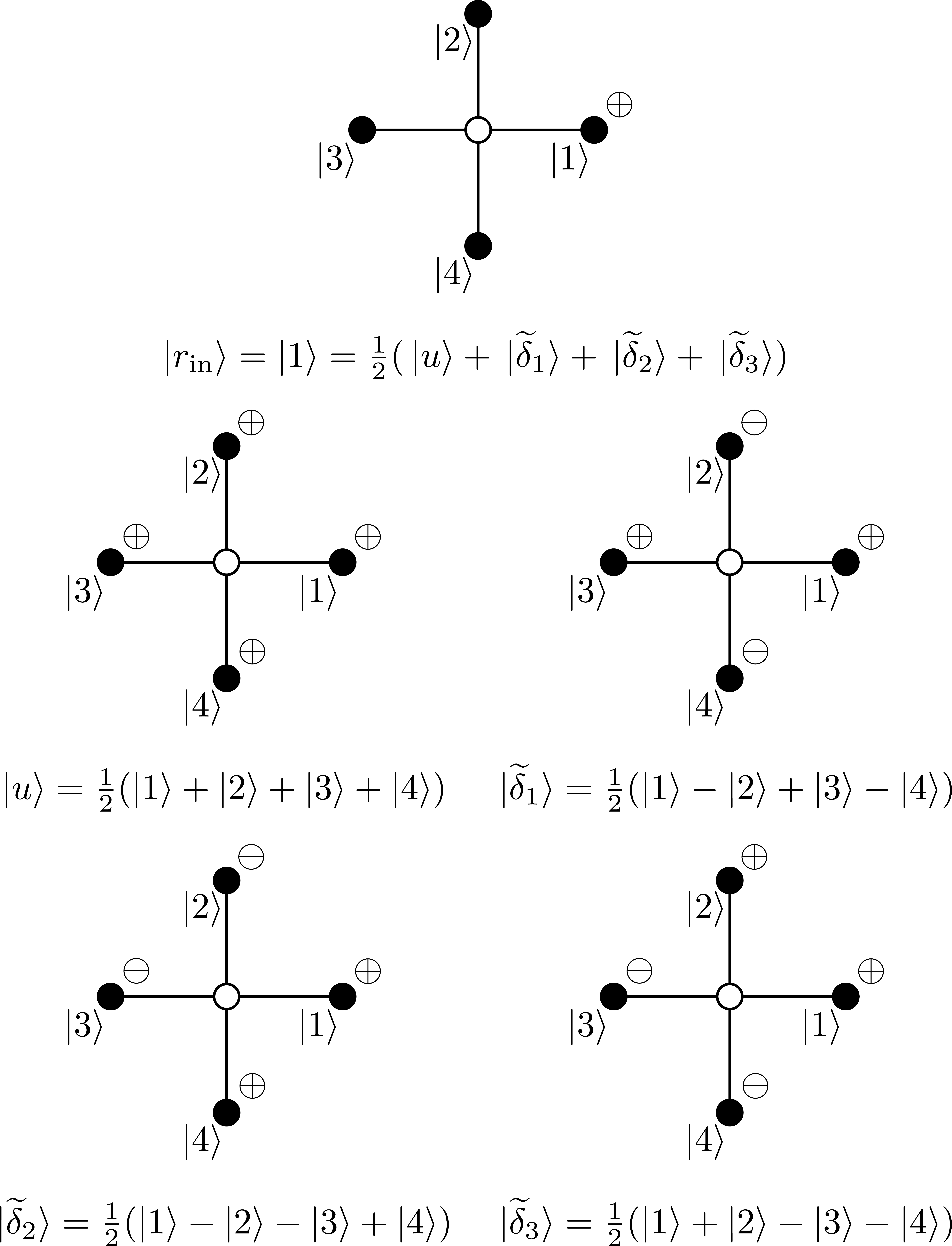}
    \caption{
      Decomposition of an initial state $\ket{1}$, localized on an exterior node of a cross, into three dark components,
      and the auxiliary uniform  state (AUS).
      Detection is attempted on the node in the center (open circle).
      We find $\nu = 4$ from the $\pi/2$ degree rotation symmetry about the detection site.
      Consequently, $\TDP(1) \le  1/4$ from the simple bound, Eq.~\eqref{eq:UpperBound}.
    }
    \label{fig3F}
  \end{figure}

  This bound can be easily generalized to other structures, for example the cube of Fig.~\ref{fig:PD}.
  Consider the transition to one vertex, denoted by $\ket{0}$, from one of its nearest neighbors, say $\ket{1}$.
  Let us denote the neighbors of $\ket{0}$ by $\ket{1}$, $\ket{2}$, and $\ket{3}$ and define an
  auxiliary state $\ket{u} := (\ket{1} + \ket{2} + \ket{3})/\sqrt{3}$.
  (This would be the generalization of $\ket{\AUS_0}$ from before.)
  Using the same procedure as before, we find that $\TDP(1) \le 1/3$.
  This trick can be easily extended and is summarized in the following proposition:

  {\em Proposition 1:}
  The total detection probability for the transition from a localized initial state $\ket{\RIn}$ to
  a localized detection state $\ket{\RDet}$ is bounded by the reciprocal of the number $\nu$ of
  nodes physically equivalent to $\RIn$:
  \begin{equation}
    \TDP(\RIn)
    \le
    \frac{1}{\nu}
    .
  \label{eq:UpperBound}
  \end{equation}

  Recall the definition of physically equivalent states from Eq.~\eqref{eq:DefPhysEq}:
  They have identical transition amplitudes to the detection state at all times.
  Such states are indistinguishable in the first detection problem.
  For a given system and a given transition $\RIn\to\RDet$ from one localized detection state to another,
  we can find a set of $\nu$ initial nodes $\{ r_j \}_{j=0}^{\nu-1}$ (where $r_0 = \RIn$) which are
  physically equivalent in this sense.
  They can be identified from elementary symmetry considerations.
  This number $\nu$ of physically equivalent states is what appears in Eq.~\eqref{eq:UpperBound}.
  The strongest bound is given by the {\em maximal} number of physically equivalent states,
  i.e. by the largest possible value for $\nu$.
  However, Eq.~\eqref{eq:UpperBound} still holds even when this maximal number can not be identified beyond doubt.

  Let's discuss the remaining examples of Fig.~\ref{fig:PD}.
  For the cube, we find $\TDP \le 1/3$ except for the diametrically opposed node, for which $\TDP \le 1$.
  For the complete graph with $L$ sites, all nodes besides the detection site are equivalent
  and we find $\TDP \le  1/(L-1)$ (here $L=8$).
  For simple-cubic lattices in $d$ dimensions with periodic boundary conditions one finds $\TDP \le 1/(2d)$
  for sites on the main horizontal, vertical and on the main diagonals, but smaller values for sites which are off these main axes.
  The Star-of-David graph with detector on one of the tips yields $\TDP \le 1/2$, due to reflection symmetry, except for the opposing tip.
  Similarly, yet less obvious $\TDP \le 1/2$ in the Tree-of-Life graph.
  For a square graph with an additional node in the center, we have $\nu=4$ for a transition from one of the
  corners to the center.
  When the detector is in one of the square's corners, the neighboring corners are equivalent yielding $\nu=2$,
  but all other sites are unique.
  Also for a tree, the number of physical equivalent sites and thus the upper bound varies with
  the position of the detector.
  The square, the ring, the complete graph, the hypercube, and the tree have been discussed above,
   where we computed $\TDP$ exactly.
  With one exception, the exact values of these examples actually coincide with the upper bound!
  The only exception is the tree, when the detector is not placed on the root node.

 Finally, the infinite line demands special attention.
  The upper bound yields $\TDP \le 1/2$ for every non-detection site and is correct.
  However, since it is an infinite system, the exact formula 
  Eq.~\ref{eq:Kessler}
does not apply.
  In particular one finds that $\TDP$ has a complicated dependence on $\tau$, see Refs.~\cite{Friedman2017b, Thiel2018a},
  where this model was investigated in detail.
  Yet, the $\TDP(\tau)$ curves stay below the upper bound $1/2$.

  To gain further physical insight, consider the cross structure presented in Fig.~\ref{fig3F}.
  We detect on the center of the cross, at node $\ket{0}$ and start on one of the outer nodes, for example on state $\ket{1}$.
  This initial state can be decomposed into a linear combination of four states, out of which three are
  easily understood as being dark states.
  For example the  state $\ket*{\tilde{\delta_1}}=(\ket{1} -\ket{2} + \ket{3} - \ket{4})/2$ is dark
  since it is not injecting probability current into state $\ket{0}$.
  Destructive interference erases all amplitude in the detection state.
  The uniform state $\ket{\AUS} =(\ket{1} + \ket{2} + \ket{3} + \ket{4})/2$
  which is a normalized sum of all the equivalent states in the system, is the fourth state.
  This state gives a constructive interference pattern at the detected state.
  Returning to the transition $\ket{1} \to \ket{0}$ we decompose the initial condition into a
  superposition of the four states, as shown in Fig. \ref{fig3F}.
  Since three components are dark, and the overlap of initial and uniform state is $1/\sqrt{4}$ we find $\TDP(1) \le 1/4$.
  In fact, a straight-forward calculation using Eq.~\eqref{eq:Kessler} shows $\TDP(u) = 1$ and thus $\TDP(1) = 1/4$.
  This is clearly in accord with $\nu = 4$.

  Let us formalize our result.
  Consider a localized initial state $\ket{\RIn} $ and assume that in
  the system we have a total of $\nu$ physically equivalent states $\{ \ket{r_j} \}_{j=0}^{\nu-1}$,
  where $\ket{r_0} = \ket{\RIn}$ and $\ip{r_i}{r_j} = \delta_{i,j}$.
  These states span the subspace $\mathcal{E}_{\ket{\RIn}} = \Span{\ket{r_j}}$, i.e. the space of all possible
  superpositions of the physically equivalent states $\ket{r_j}$.
  Let
  \begin{equation}
    \ket*{\AUS(\RIn)}
    :=
    \frac{1}{\sqrt{\nu}} \Sum{j=0}{\nu-1} \ket{r_j}
  \label{eq:DefAUS}
  \end{equation}
  be the {\em auxiliary uniform state} (AUS).
  $\ket{\AUS(\RIn)}$ is one particular state in the subspace $\mathcal{E}_{\ket{\RIn}}$.
  We can find a new basis $\{ \ket{\AUS(\RIn)}, \ket*{\overline{\delta}_1}, \hdots, \ket*{\overline{\delta}_{\nu-1}} \}$
  for $\mathcal{E}_{\ket{\RIn}}$, that consists of the AUS and $\nu-1$ dark states.
  These dark states can be found similarly to the stationary dark states in
Sec.  
\ref{sec:Dichotomy}.
  The solution reads:
  \begin{equation}
    \ket*{\overline{\delta}_j}
    =
    \frac{j \ket{r_{j}} - \sSum{m=0}{j-1} \ket{r_m}}{\sqrt{j (j+1) } }
    ,
  \label{eq:DarkPhysEqSol}
  \end{equation}
  where $j=1,2,\hdots, \nu-1$.
  One quickly verifies that the states $\ket*{\overline{\delta}_j}$ are normalized, orthogonal to each other and to the AUS.
  Furthermore they are dark, because the transition amplitudes to the detection state from any state $\ket{r_j}$ are the same.
  Consequently, the multiplication with $\bra{\RDet}\TEO(t)$ annihilates $\ket*{\overline{\delta}_j}$, because it yields
  $\mel*{\RDet}{\TEO(t)}{\overline{\delta}_j} = \mel*{\RDet}{\TEO(t)}{\RIn} [ j - j ]/ \sqrt{j(j+1)} = 0$.
  The original initial state $\ket{\RIn} = \ket{r_0}$ is now rewritten as
  \begin{equation}
    \ket{\RIn}
    =
    \frac{1}{\sqrt{\nu}} \ket*{\AUS(\RIn)}
    - \Sum{j=1}{\nu-1} \frac{\ket*{\overline{\delta}_j}}{\sqrt{j(j+1)}}
    .
  \label{eq:}
  \end{equation}
 Knowing about the dark states in $\ket{\RIn}$, we can directly obtain $\TDP(\RIn)$,
  because $\FDA_n(\overline{\delta}_j) = 0$.
  We immediately find:
  \begin{equation}
    \TDP(\RIn)
    =
    \frac{1}{\nu} \TDP(u(\RIn))
    .
  \label{eq:AUSAndNu}
  \end{equation}
  Since $\TDP(\AUS) \le 1$, we obtain the upper bound Eq.~\eqref{eq:UpperBound}.
  It is important to keep in mind that $\nu$ is not a global property of the system,
  but depends on the transition $\RIn\to\RDet$.
  Clearly, if the AUS is bright, then $\TDP(\AUS(\RIn)) = 1$, and the upper bound saturates,
  i.e. it becomes an equality.
  We also note, that Eq.~\eqref{eq:AUSAndNu} holds as well in infinite systems.
Some of us are currently studying a group theoretic approach to the problem,
which can extend the symmetry considerations presented here. 
A lower bound was also recently presented, including the discussion
of hypercube in dimension $d$, i.e. the extension of this work
to certain large systems with what we called a shell structure 
 was considered \cite{Thiel2020a}.

\section{Summary}
\label{sec:Sum}
  We have investigated herein the total probability of detection, $\TDP$, in a quantum system that is stroboscopically probed in its detection state. An explicit formula for this in terms of the energy eigenstates was produced via the renewal equation \cite{Gruenbaum2013a,Friedman2017b}
previously derived for the generating function for the detection amplitude, with the help of the Aleksandrov theorem for Cauchy transforms. An alternate derivation for the formula was also given via an analysis of the dark and bright subspaces that comprise 
  the total Hilbert space. 
  The dark states are those energy eigenstates that have no overlap with the detector and thus are never detected. 
  There were found to be  two classes of dark states: those that belong to completely dark energy levels, where every eigenstate of the sector is orthogonal to the detection state, and those which perforce appear in degenerate energy levels, whose sectors possess a non-zero projection of the detection state.
  The bright states are those eigenstates which are detected with probability unity, were shown to constitute, in a finite system, the orthogonal complement to the dark subspace. They were shown to  belong to the spectrum of the survival operator $\Surf$ inside the unit disk. An explicit set of basis states for the subspaces was constructed.  From this, $\TDP$ was calculated as the  overlap of the initial state with the bright space, reproducing the original result.
  We considered several examples, showing in particular how lifting the degeneracy in the energy spectrum discontinuously changes $\TDP$. The breakdown of our formula for $\TDP$ in an infinite system was discussed in the context of the infinite line.

  The $\tau$-independence of the total detection probability will survive when irregularities are 
  introduced in the sampling times.
  The lack of the total control over the detector can be modeled by a random sequence of inter-detection times $\{\tau_1, \tau_2, \hdots \}$,
  as has been done, e.g. in Ref.~\cite{Varbanov2008a}, where the sampling times were given by a Poisson process.
  A strong finding of ours is that $\TDP$ does actually not depend on $\tau$ at all for systems with a discrete spectrum.
  As the set of resonant detection periods [defined by Eq.~\eqref{eq:DefResonantTau}] has zero measure,
  and thus all of our results are expected to hold for non-stroboscopic sampling as well.

  We also investigated the influence of the system's symmetries on $\TDP$.
  Whenever one has found $\nu$ physically equivalent initial states, one will find that
  the total detection probability is bounded by $\TDP \le 1/\nu$.
  Two states are physically equivalent when they yield the same transition amplitudes to the detection
  state for all times.
  Any pair of equivalent states can be seen to yield a dark state from their negative superposition.
  The AUS defined by Eq.~\eqref{eq:DefAUS} is the positive superposition of all equivalent states, contains all bright components of the
  original initial state, and gives $\TDP$ via Eq.~\eqref{eq:AUSAndNu}.

  \acknowledgements
    The support of Israel Science Foundation's grant 1898/17 is acknowledged.
    FT is supported by DFG (Germany) under grant TH 2192/1-1.
    EB thanks Avihai Didi for comments on the manuscript and Klaus Ziegler for fruitful discussions.

\appendix
\renewcommand\thefigure{\thesection\arabic{figure}} 
\setcounter{figure}{0}  
 \section{Aleksandrov's theorem in a two-level system\label{AppTwoLev}}
    A proof of the powerful Eq.~\eqref{eq:Aleksandrov} lies far outside the scope of this 
    paper, the interested reader is referred to \cite{Cima2006a}.
    In order to nevertheless help to understand the origin of the equality, we consider the case when there are only two different 
    energy levels, $E_1$ and $E_2$.
    The Hamiltonian of such a system would read $\Ham = \hat{P}_1 E_1 + \hat{P}_2 E_2$.
    It is not necessary to specify the eigenstates and degeneracies, as long as the overlap of each 
    eigenspace with the detection state is given.
    For simplicity, we assume that $\ev*{\hat{P}_1}{\PsiDet} = \ev*{\hat{P}_2}{\PsiDet} = 1/2$.
    (In fact, the Aleksandrov ansatz does not ``see'' individual eigenstates $\ket{E_{l,m}}$, 
    but just the overlap with the eigenspace projectors $\hat{P}_l$.)
    We can replace $\mel*{\PsiDet}{\hat{P}_1}{\PsiIn} = \nu(\lambda_1) \ev*{\hat{P}_1}{\PsiDet}%
    = \nu(\lambda_1) /2$, and similarly with $\nu(\lambda_2)$.
    $\FDA(z)$ reads:
    \begin{equation}
      \FDA(z)
      =
      \frac{
          \frac{
            z e^{-i\lambda_1} \nu(\lambda_1)
          }{
            1 - z e^{-i \lambda_1}
          }
          +
          \frac{
            z e^{-i\lambda_2} nu(\lambda_2)
          }{
            1 - z e^{-i \lambda_2}
          }
      }{
        \frac{
          1
        }{
          1 - z e^{- i \lambda_1}
        }
        +
        \frac{
          1
        }{
          1 - z e^{- i \lambda_2}
        }
      }
      .
    \label{eq:FDATwo}
    \end{equation}
    Its absolute value on the unit circle equals:
      \begin{align}
        \abs*{\FDA(e^{i\theta})}^2
        = \nonumber 
        \frac{
          \nu(\lambda_1)^* e^{i\lambda_1} ( e^{i\lambda_2} - e^{i\theta} )
          +
          \nu(\lambda_2)^* e^{i\lambda_2} ( e^{i\lambda_1} - e^{i\theta} )
        }{
          e^{i\lambda_1} + e^{i\lambda_2}
          - 2 e^{i\theta}
        }
        \\ \times
        \frac{
          \nu(\lambda_1) e^{-i\lambda_1} ( 1 - e^{i\theta - i \lambda_2} )
          +
          \nu(\lambda_2) e^{-i\lambda_2} ( 1 - e^{i\theta - i \lambda_1} )
        }{
          2 - e^{i\theta}
          ( e^{-i\lambda_1} + e^{-i\lambda_2} )
        }
      \label{eq:AbsValueExample}
      \end{align}
      When this expression is plugged into Eq.~\eqref{eq:Aleksandrov}, the left-hand side integral can 
      be treated by a variable change $z = e^{i\theta}$, and $\dd \theta = \dd z / (i z)$.
      The result is a complex contour integral which is solved by residue inspection.
      The integrand has only two simple poles inside the unit circle.
      One lies at the origin $z=0$ and the other is defined by the linear term in the denominator of the 
      first line in Eq.~\eqref{eq:AbsValueExample}.
      After some lengthy algebra, we find that the prefactors of the cross-terms $\nu(\lambda_1)^* \nu(\lambda_2)$ and $\nu(\lambda_2)^* \nu(\lambda_1)$ vanish.
      Collecting the residues of the diagonal terms, however, we find one half:
      \begin{equation}
        \oint\limits_{\abs{z}=1} \frac{\dd z}{2\pi i z}
        \frac{
          e^{i\lambda_2}
          ( 1 - z e^{-i\lambda_2} )
        }{
          \qty[
            e^{i\lambda_1} + e^{i\lambda_2}
            - 2 z
          ] \qty[
            2 - z
            ( e^{-i\lambda_1} + e^{-i\lambda_2} )
          ]
        }
        =
        \frac{1}{2}
        .
      \label{eq:TwoState}
      \end{equation}
      We obtain the same result for the integral proportional to $\abs{\nu(\lambda_2)}^2$.
      Hence: 
      \begin{align}
        \TDP
        = & 
        \frac{1}{2} \qty[
          \abs{\nu(\lambda_1)}^2 
          + \abs{\nu(\lambda_2)}^2 
        ]
        ,
      \label{eq:TDPTwo}
      \end{align}
      which is the result of Eq.~\eqref{eq:Aleksandrov} and Eq.~\eqref{eq:Kessler} for the chosen example system.
      A demonstration with more than two energy levels using the same method first becomes tedious and soon unfeasible.
      In a system with $w$ bright states there are $w$ different phases $\lambda_l$ in $\mu(\theta)$ and $w$ poles in the integrand of Eq.~\eqref{eq:TDPHardy}
      after switching again to $z = e^{i\theta}$.
      Aleksandrov's theorem ``magically'' ensures that the residues of these poles are exactly given by $\abs{\nu(\lambda_l)}^2 \ev*{\hat{P}_l}{\PsiDet}$ 
      thus giving Eq.~\eqref{eq:Kessler}.

\section{Additional properties of the Survival operator}
\label{app:Surf}
  Here, we will give a more detailed discussion of $\Surf$'s properties.
  In the main text we already mentioned that the stationary dark states are eigenstates of $\Surf$ with 
  eigenvalue on the unit circle and that the stationary bright states belong to the spectrum inside the unit disk.
  In view of the fact that $\Surf$ is not a diagonalizable matrix in general, these statements need to be refined.

  First, we note that the stationary dark states $\ket{\delta_{l,m}}$ of Eq.~\eqref{eq:StationaryDark} are in general
  the right-eigenstates of $\Surf$ that correspond to eigenvalues $e^{- i \tau E_l / \hbar}$ on the unit circle.
  In particular, these eigenvalues coincide with eigenvalues of $\TEO(\tau)$.
  Next, Eq. \eqref{eq:Determ} that relates the poles of the generating function  $\FDA(z)$ 
  and the spectrum of $\Surf$ is generally true,  as well.
  Its algebraic eigenvalues $\zeta$ are given by the zeros of the characteristic polynomial $\determinant[ \zeta \Id - \Surf ]$.

  It remains to be shown that the survival probability $S_n(\tilde{\beta}) = \norm*{\Surf^n\ket*{\tilde{\beta}}}$ of some bright state 
  decays to zero as $n\to\infty$.
  To see this, consider the restriction of $\Surf$ to the bright space $\Surf_B = \PBright \Surf \PBright$.
  This removes all trivial eigenvalues that lie on the unit circle from $\Surf$.
  This operator is defective in general and can not be brought into a diagonal form.
  It can be brought into a Jordan normal form, though:
  \begin{equation}
    \Surf_B
    =
    \Sum{l}{} 
    \sbrr{ \zeta_l \hat{M}_l + \hat{N}_l }
    .
  \label{eq:SurvBright}
  \end{equation}
  Here $\zeta_l$ are the algebraic eigenvalues of $\Surf_B$ that we determined from $\FDA(z)$,
  and the matrices $\hat{M}_l$ and $\hat{N}_l$ act on the (generalized) eigenspaces of $\zeta_l$.
  $\hat{N}_l$ is a nilpotent matrix (i.e. $\hat{N}_l^{p_l} = 0$, for some integer $p_l>0$) and $\hat{M}_l$ is 
  the projector to the corresponding eigenspace (hence $\hat{M}_l^2 = \hat{M}_l$).
  The rank of $\hat{M}_l$ is equal to the algebraic multiplicity of $\zeta_l$,
  $\hat{M}_l$ and $\hat{N}_l$ commute, and $\hat{M}_l \hat{N}_{l'} = \hat{N}_{l'} \hat{M}_l = \delta_{l,l'} \hat{N}_{l'}$.
  Using these properties, one finds for $n>p_l$
  \begin{equation}
    (\zeta_l \hat{M}_l + \hat{N}_l )^n
    =
    \zeta_l^n \hat{M}_l
    +
    \Sum{m=1}{p_l-1}
    \binom{n}{m}
    \zeta_l^{n-m}
    \hat{N}_l^m
    ,
  \label{eq:NilPotent}
  \end{equation}
  and therefore for any superposition $\ket*{\tilde{\beta}}$ of bright states:
  \begin{equation}
    \SP_n(\tilde{\beta)}
    =
    \norm*{\Surf_B^n \ket*{\tilde{\beta}}}^2
    =
    \Sum{l}{} \abs{\zeta_l}^{2n} 
    C_{l,n}(\tilde{\beta})
    .
  \label{eq:SPSuper}
  \end{equation}
  The coefficients are given by:
  \begin{equation}
    C_{l,n}(\tilde{\beta})
    :=
    \norm{ [\hat{M}_l + \zeta_l^{-1} \hat{N}_l ]^n \ket*{\tilde{\beta}}}^2
    .
  \label{eq:NormalCoeff}
  \end{equation}
  Using Eq.~\eqref{eq:NilPotent}, the triangle inequality, as well as $\abs{\zeta_l}^{-m} < \abs{\zeta_l}^{-p_l}$ and 
  $\binom{n}{m} \le \binom{n}{p_l} \le n^{p_l} / p_l!$, for $n> 2p_l$ and $m<p_l$, one can bound these coefficients with:
  \begin{align}
    C_{l,n}(\tilde{\beta})
    \le & \nonumber
    \norm*{\hat{M}_l\ket*{\tilde{\beta}}}^2
    +
    \Sum{m=1}{p_l-1} \binom{n}{m}^2 \abs{\zeta_l}^{-2m}
    \norm*{\hat{N}_l^m\ket*{\tilde{\beta}}}^2
    \\ \le & 
    \norm*{\hat{M}_l\ket*{\tilde{\beta}}}^2
    +
    \frac{( n \abs{\zeta_l^{-1}})^{2p_l}}{p_l!^2}
    \Sum{m=1}{p_l-1} 
    \norm*{\hat{N}_l^m\ket*{\tilde{\beta}}}^2
    .
  \end{align}
  Hence $C_{l,n}(\tilde{\beta}) \le \sLandau{n^{2p_l}}$.
  Considering the largest of all $p_l$ denoted by $p_*$ and the eigenvalue $\zeta_*$ closest to the unit circle,
  we find that the survival probability will eventually decay exponentially
  \begin{equation}
    \SP_n(\tilde{\beta})
    \le \Landau{ n^{2p_*} \abs{\zeta_*}^{2n}}
    .
  \label{eq:SPdecay}
  \end{equation}

  From writing $\sbrr{\Id - \Detect}$ in the energy basis and inspecting the definition of $\Surf_B$,
  it is evident that each $\zeta_l$ has to be a {\em convex}
  sum of the phase factors $e^{i\tau E_l/\hbar}$.
  In fact, Ref.~\cite{Gruenbaum2013a} shows that the non-trivial eigenvalues $\zeta_l$ can be obtained 
  from the stationary points of a certain 2d-Coulomb force fields and must lie inside the unit disk.
  This approach is investigated in detail in Ref.~\cite{Yin2019a}.
  Therefore we find another reason why the non-trivial eigenvalues obey $\abs{\zeta_l} < 1$.

  This is easily seen for a system with only two bright states $\ket{\beta_1}$ and $\ket{\beta_2}$ 
  corresponding to two energies $E_1$ and $E_2$.
  The detection state is decomposed into the bright states via $\ket{\PsiDet} = a \ket{\beta_1} + b\ket{\beta_2}$,
  where $\abs{a}^2 + \abs{b}^2 = 1$.
  In this case, $\Surf_B$ reads in a matrix notation:
  \begin{align}
    \Surf_B
    = &
    \mqty( 1 - \abs{a}^2 & a b^* \\ b a^* & 1 - \abs{b}^2 )
    \mqty( e^{-i \frac{\tau E_1}{\hbar}} & 0 \\ 0 & e^{-i \frac{\tau E_2}{\hbar}} )
    \\ = &
    \mqty( \abs{b}^2 e^{-i \frac{\tau E_1}{\hbar}} & a b^* e^{-i \frac{\tau E_2}{\hbar}} \\ 
        b a^* e^{- i \frac{\tau E_1}{\hbar}} & \abs{a}^2 e^{- i \frac{\tau E_2}{\hbar}} 
    )
    .
  \label{eq:SurfTwo}
  \end{align}
  This matrix has one vanishing eigenvalue $\zeta_1  = 0$.
  The other one is equal to $ \zeta_2 = \abs{a}^2 e^{- i \tau E_2 / \hbar} + \abs{b}^2 e^{- i \tau E_1 /\hbar}$, 
  i.e. it is a convex sum of the phase factors, which lies inside the unit circle.
  Similar reasoning applies to systems with more than two bright states.

  In systems with finite dimensional Hilbert space we are finished, because there is some minimum 
  decay rate $\lambda^* = \min_l\sofff{-2\ln\abs{\zeta_l}} > 0$.
  Infinitely dimensional systems may behave more subtly.
  In the thermodynamic limit the eigenvalues $\zeta_l$ can get 
  infinitely close to the unit circle.
  In that case unity is an accumulation point of the sequence $\{\abs{\zeta_l}\}$ and $\lambda^* = \inf_l\{-2\ln\abs{\zeta_l}\} = 0$.
  In the view of the spectral theorem of Ref.~\cite{Gruenbaum2013a}, there are two options:
  Either, one is lucky and $\SP_n$ decays to zero, albeit slower than exponentially fast, or not.
  $\SP_n\to0$ will still hold in the first case.
  The spectral theorem of Ref.~\cite{Gruenbaum2013a} says, that this is the case when the spectrum of $\TEO(\tau)$ has no absolutely continuous part.
  If such a part of the spectrum is present, on the other hand, then $\SP_n$ will converge 
  to some positive value, and we find $\TDP(\tilde{\beta}) < 1$.
  This is obviously only possible in infinitely dimensional systems.
  Some more mathematical details about $\Surf_B$ can be found in 
  Appendix A of Ref.~\cite{Bourgain2014a}.

\section{Additional Example Graphs and their $\TDP$ and dark states\label{sec:ExamplesII}}
In this appendix, we present the calculation for the spectrum, eigenstates, $\TDP$ and dark states for the collection of graphs illustrated in Fig. \ref{fig:ExampleGraphs2}(C-F),  (A) and (B) having been presented in the main text.

  \begin{figure*}
    \centering
    \includegraphics[width=0.99\textwidth]{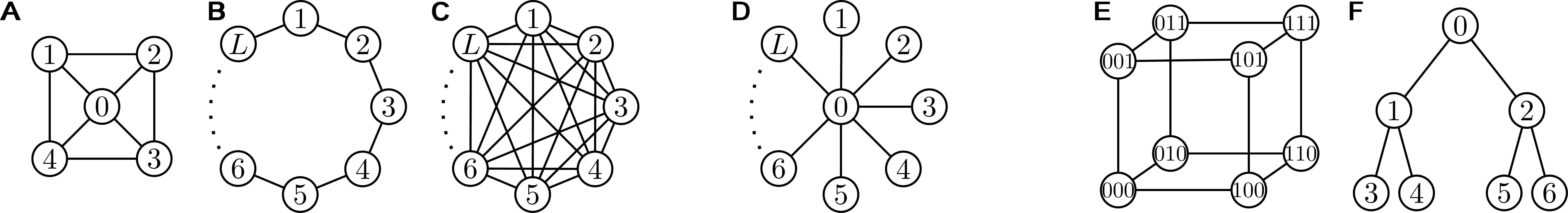}
    \caption{
      Notation used for the examples.
      The letters also correspond to the subsections in Sec.~\ref{sec:Examples}.
      (A) Square with a center node,
      (B) ring of $L$ sites,
      (C) complete graph with $L$ sites,
      (D) star graph with $L$ sites in the periphery
      (E) hypercube (here of dimension $d=3$
      (F) binary tree with $2$ generations.
      \label{fig:ExampleGraphs1}
    }
  \end{figure*}

 \subsection{Complete graph}
    We  now consider a graph with $L$ nodes, in which each node is connected to each other node, see \ref{fig:ExampleGraphs1}C.
    Its Hamiltonian reads:
    \begin{equation}
      \Ham
      =
      -\gamma 
      \Sum{r,r'=1}{L} (1 - \delta_{r,r'}) \dyad{r}{r'}
    \label{eq:CompleteH}
    \end{equation}
    We pick one node as the detection node, say $\RDet$.

    The system has only two energy levels, namely $E_1 = -\gamma(L-1)$ and $E_2 = \gamma$,
    the latter being $(L-1)$-fold degenerate.
    The eigenstates are:
    \begin{align}
      \ket{E_1} = \ket{j} := \Sum{r=1}{L} \frac{\ket{r}}{\sqrt{L}}
      \qc
      \ket{E_{2,m}} = \Sum{r=1}{L} \frac{e^{i \frac{2\pi mr}{L}} \ket{r}}{\sqrt{L}}
    \end{align}
    where $m=1,2,\hdots,L-1$ and $\ket{j}$ is the uniform state over all nodes.
    Therefore, none of these energy levels is completely dark with respect to a localized detection state.
    The two bright eigenstates are:
    \begin{equation}
      \ket{\beta_1} = \ket{j} = \frac{1}{\sqrt{L}} \Sum{r=1}{L} \ket{r}
      \qc
      \ket{\beta_2} = \frac{\sqrt{L} \ket{\RDet} - \ket{j}}{\sqrt{L-1}}
      .
    \label{eq:CompleteBright}
    \end{equation}
    Computing the overlap with the bright space, we find:
    \begin{equation}
      \TDP(\RIn)
      =
      \left\{ \begin{aligned}
        1\qc & \RIn = \RDet \\
        \tfrac{1}{L-1} \qc & \RIn \ne \RDet
      \end{aligned} \right.
      .
    \label{eq:TDPComplete}
    \end{equation}

  \subsection{Star}
    The next example is a star graph with a center node $\ket{0}$ and $L$ nodes in the periphery, see Fig.~\ref{fig:ExampleGraphs1}D.
    The Hamiltonian reads:
    \begin{equation}
      \Ham
      =
      - \gamma \Sum{r=1}{L} \qty[
        \dyad{0}{r} + \dyad{r}{0}
      ]
      .
    \label{eq:StarH}
    \end{equation}
    The system has three energy levels $E_1 = - \gamma \sqrt{L}$, $E_2 = 0$ and $E_3 = \gamma\sqrt{L}$, 
    of which $E_2$ is $(L-1)$-fold degenerate.
    The eigenstates are:
    \begin{align}
      \left\{ \begin{aligned}
        \ket{E_1} = \frac{\ket{0} + \ket{j}}{\sqrt{2}}
        \qc
        \ket{E_3} = \frac{\ket{0} - \ket{j}}{\sqrt{2}}
        \\ 
        \ket{E_{2,m}} = \frac{1}{\sqrt{L}} \Sum{r=1}{L}
        e^{i \frac{2\pi}{L} m r} \ket{r}
      \end{aligned} \right.
      ,
    \end{align}
    where $m=1,2,\hdots, L-1$. 
    $\ket{j} := \sSum{r=1}{L}\ket{r}/\sqrt{L}$ is the uniform state over the periphery.
    
    If the detection takes place on the center node, the energy level $E_2$ is completely dark.
    Would detection take place in the periphery, there would be no completely dark energy levels.
    $\TDP$ is computed from Eq.~\eqref{eq:Kessler}.
    The calculations are similar to the ones of the complete graph:
    \begin{equation}
      \TDP(\RIn)
      =
      \left\{ \begin{aligned}
        1
        \qc & \RDet = \RIn \\
        \tfrac{1}{L} 
        \qc & \RDet = 0, \RIn \ne 0 \\
        1
        \qc & \RDet \ne 0, \RIn = 0 \\
        \tfrac{1}{L-1}
        \qc & \RDet \ne 0, \RIn \ne 0
      \end{aligned} \right.
      .
    \label{eq:TDPStar}
    \end{equation}

  \subsection{Hypercube}
    The next example is the $d$-dimensional hypercube.
    An example graph for $d=3$ is depicted in Fig.~\ref{fig:ExampleGraphs1}E.
    In this system there are $2^d$ different states which can be represented by a $d$-length string 
    consisting only of zeros and ones, i.e. a string of $d$ bits, $\ket{\psi} = \otimes_{j=1}^d \ket{b_j}$,
    where $b_j = 0_j, 1_j$ is the $j$-th bit.
    The allowed transitions correspond to bit flips, and the Hamiltonian can be written as a sum of Pauli $\hat{\sigma}_x$-matrices:
    \begin{equation}
      \Ham = - \gamma \Sum{j=1}{d} \hat{\sigma}_x^{(j)}
      ,
    \label{eq:HyperH}
    \end{equation}
    where $\hat{\sigma}_x^{(j)}$ only acts on $\ket{b_j}$, and $\sigma_x^{(j)} = \dyad{0_j}{1_j} + \dyad{1_j}{0_j}$.
    (As an example: $\hat{\sigma}_x^{(1)} \ket{110} = \ket{010}$, $\hat{\sigma}_x^{(2)} \ket{110} = \ket{100}$, and $\hat{\sigma}_x^{(3)} \ket{110} = \ket{111}$.)
    The relation to a word of $d$ bits with bit-flip transitions make this model particularly relevant 
    for the quantum computation field \cite{Kempe2005a, Krovi2006a}.
    The energy levels are $E_l = -\gamma (2l-d)$, $l=0,1,\hdots,d$, each is $\binom{d}{l}$-fold degenerate.
    Let $x_j = \pm 1$ and define $\ket{x_j} := (\ket{0_j} + x_j\ket{1_j})/\sqrt{2}$, then $\hat{\sigma}^{(j)}_x\ket{x_j} = x_j \ket{x_j}$.
    Hence the energy eigenstates have the form:
    \begin{equation}
      \ket{E_{l,m}}
      =
      \otimes_{j=1}^{d} \ket{x_j}
      ,
    \label{eq:HyperE}
    \end{equation}
    such that $\sSum{j=1}{d} x_j = 2l-d$ and $m$ enumerates all $\binom{d}{l}$ combinations.
    In each bit, we have the important relation:
    \begin{equation}
      \ip{b_j}{x_j} = \frac{1}{\sqrt{2}} (-1)^{b_j \frac{1-x_j}{2}} 
        ,
    \label{eq:BitOverlap}
    \end{equation}
    which has a negative sign when $b_j = 1$ and $x_j = -1$ and a positive sign otherwise.
    
    We try to detect the system in the node with all bits equal to zero, i.e. $\ket{\RDet} = \otimes_j \ket{0_j}$.
    From Eq.~\eqref{eq:BitOverlap} we find that all energy eigenstates have the same overlap with the detection state: $\ip*{\RDet}{E_{l,m}} = 2^{-d/2}$.
    Therefore one finds the following bright eigenstates:
    \begin{equation}
      \ket{\beta_l}
      =
      \frac{1}{\sqrt{\binom{d}{l}}}
      \Sum{ \abs*{\{x_j\}} = 2l-d}{} \otimes_{j=1}^d \ket{x_j}
      ,
    \label{eq:StatBrightHyperCube}
    \end{equation}
    where the sum runs over all combinations $\{ x_j \}$, where $\abs*{\{x_j\}} = \sSum{j=1}{d} x_j = 2l-d$.

    We now pick any localized initial state $\ket{\RIn} = \otimes_{j=1}^d \ket{b_{\text{in},j}}$, which differs from $\ket{\RDet}$ in exactly $\xi$ bits.
    ($\xi$ is the so-called Hamming-distance between both states.)
    $\TDP$ only depends on $\xi$ and can be obtained from Eq.~\eqref{eq:Kessler}, and Eq.~\eqref{eq:StatBrightHyperCube} and some complicated combinatoric computation.
    We do not present this here, and instead refer to Ref.~\cite{Novo2015a}, where $\TDP$ was computed in a different way:
    \begin{equation}
      \TDP(\RIn)
      =
      \frac{1}{\binom{d}{\xi}}
      .
    \label{eq:TDPHyper}
    \end{equation}
    This shows that only two transitions in a hypercube are actually reliable, in the sense that $\TDP =1$: 
    the return to the initial node and the traversal to the opposing node, see Fig.~\ref{fig:ExampleGraphs2}E.

  \subsection{A tree graph}
    Similar to the hypercube, trees are important examples from quantum computation,
    in particular because they feature an exponential speedup \cite{Farhi1998a, Childs2002a} relative to classical algorithms.

    We will consider here only a small binary tree with two generations, see Fig.~\ref{fig:ExampleGraphs1}F.
    The top of the tree is called the root, the bottom nodes are called leaves of the tree.
    Its Hamiltonian reads:
    \begin{equation}
      \Ham
      =
      - \gamma
      \mqty(
        0 & 1 & 1 & 0 & 0 & 0 & 0 \\
        1 & 0 & 0 & 1 & 1 & 0 & 0 \\
        1 & 0 & 0 & 0 & 0 & 1 & 1 \\
        0 & 1 & 0 & 0 & 0 & 0 & 0 \\
        0 & 1 & 0 & 0 & 0 & 0 & 0 \\
        0 & 0 & 1 & 0 & 0 & 0 & 0 \\
        0 & 0 & 1 & 0 & 0 & 0 & 0 \\
      )
      .
    \label{eq:Tree}
    \end{equation}
    The corresponding energy levels are: $E_1 = -2\gamma$, $E_2 = -\sqrt{2}\gamma$, $E_3 = 0$, being thrice degenerate, as well as, $E_4 = \sqrt{2}\gamma$ and $E_5 = 2\gamma$.
    The energy eigenstates are 
    \begin{align}
      \left\{ \begin{aligned}
        \ket{E_1}     = & (2,2,2,1,1,1,1)^T/4 \\
        \ket{E_2}     = & (0,-\sqrt{2},\sqrt{2},-1,-1,1,1)^T/\sqrt{8} \\
        \ket{E_{3,1}} = & (0,0,0,1,-1,0,0)^T/\sqrt{2} \\
        \ket{E_{3,2}} = & (0,0,0,0,0,1,-1)^T/\sqrt{2} \\
        \ket{E_{3,3}} = & (-2,0,0,1,1,1,1)^T/\sqrt{8} \\
        \ket{E_4}     = & (0,\sqrt{2},-\sqrt{2},-1,-1,1,1)^T/\sqrt{8} \\
        \ket{E_5}     = & (2,-2,-2,1,1,1,1)^T/4
      \end{aligned} \right.  
      .
    \end{align}
    We see that some of the energy levels may be completely dark depending on the detection state.

    \paragraph{Detection on the root.}
      The choice $\ket{\RDet} = \ket{0}$ renders $E_2$ and $E_4$ completely dark.
      The third energy level yields the bright eigenstate $\ket{\beta_3} = \ket{E_{3,3}}$.
      Doing the calculations with Eq.~\eqref{eq:Kessler}, we find the total detection probability equal to $1/2$ for nodes $\ket{1}$ and $\ket{2}$ 
      and equal to $1/4$ for the leaves, i.e. 
      \begin{equation}
        \TDP(\RIn) 
        =
        \left\{ \begin{aligned}
          1\qc & \RIn = 0 \\
          \tfrac{1}{2} \qc & \RIn = 1,2 \\
          \tfrac{1}{4} \qc & \text{otherwise}
        \end{aligned} \right.
        .
      \label{eq:TDProot}
      \end{equation}

    \paragraph{Detection in the middle.}
      Now, we choose $\ket{\RDet} = \ket{1}$.
      Then $E_3$ is completely dark.
      All remaining energy levels are non-degenerate, hence there are no other dark states.
      We find from Eq.~\eqref{eq:Kessler}:
      \begin{equation}
        \TDP(\RIn)
        =
        \left\{ \begin{aligned}
          \tfrac{1}{2} \qc & \RIn = 0 \\
          1 \qc & \RIn = 1,2 \\
          \tfrac{3}{8}\qc & \text{otherwise}
        \end{aligned} \right.
      \label{eq:TDPMiddle}
      \end{equation}

    \paragraph{Detection in the leaves.}
      We choose $\ket{\RDet} = \ket{3}$, then there are two dark state in the energy level $E_3$.
      The bright state in this level is 
      \begin{equation}
        \ket{\beta_3}
        =
        \frac{2\ket{0} - 5 \ket{3} + 3 \ket{4} - \ket{5} - \ket{6}}{\sqrt{40}}
        .
      \label{eq:LeavesBright}
      \end{equation}
      The remaining bright states are equal to the eigenstates.
      This results in 
      \begin{equation}
        \TDP(\RIn)
        =
        \left\{ \begin{aligned}
          \tfrac{3}{5}\qc & \RIn = 0,4 \\
          1\qc & \RIn=1,2,3 \\
          \tfrac{2}{5}\qc & \RIn=5,6
        \end{aligned} \right.
      \label{eq:TDPLeaves}
      \end{equation}

      All these results are  summarized in Fig.~\ref{fig:ExampleGraphs2}.

\end{document}